\documentclass[fleqn,usenatbib]{mnras}

\usepackage{newtxtext,newtxmath}
\usepackage{pdflscape}	
\usepackage{multirow}
\usepackage{makecell}
\usepackage{url}
\usepackage[T1]{fontenc}

\DeclareRobustCommand{\VAN}[3]{#2}
\let\VANthebibliography\thebibliography
\def\thebibliography{\DeclareRobustCommand{\VAN}[3]{##3}\VANthebibliography}

\usepackage{graphicx}	
\usepackage{amsmath}	

\title[Short title, max. 45 characters]{Predicting Astrometric Microlensing Events from Gaia DR3}

\author[Jie Su et al.]{
Jie Su,$^{1,2,3,4}$\thanks{E-mail: sujie1@ynao.ac.cn}
Jiancheng Wang,$^{1,2,3}$
Yigong Zhang$^{5}$\thanks{E-mail: kmuzhyg@163.com}
Xiangming Cheng$^{1,2,3,4}$
and Lei Yang$^{1,2,3}$
\\
$^{1}$Yunnan Observatories, Chinese Academy of Sciences,  Kunming 650216, China\\
$^{2}$Key Laboratory for the Structure and Evolution of Celestial Objects,Chinese Academy of Sciences,  Kunming 650216, China\\
$^{3}$Center for Astronomical Mega-Science, Chinese Academy of Sciences, Beijing 100012, China\\
$^{4}$ University of Chinese Academy of Sciences, Beijing 100049, China\\
$^{5}$ College of Information Engineering, Kunming University, Kunming 650214, China
}

\date{Accepted XXX. Received YYY; in original form ZZZ}

\pubyear{2015}

\begin{document}
\label{firstpage}
\pagerange{\pageref{firstpage}--\pageref{lastpage}}
\maketitle

\begin{abstract}
Currently astrometric microlensing is the only tool that can directly measure the mass of a single star, it can also help us to detect compact objects like isolated neutron stars and black holes. The number of microlensing events that are being predicted and reported is increasing. In the paper, the potential lens stars are selected from three types of stars, high-proper-motion stars, nearby stars and high-mass stars. For each potential lens star, we select a larger search scope to find possible matching sources to avoid missing events as much as possible. Using Gaia DR3 data, we predict 4500 astrometric microlensing events with $\delta \theta_+>0.1 mas$ that occur between J2010.0 and J2070.0, where 1664 events are different from those found previously. There are 293 lens stars that can cause two or more events, where 5 lens stars can cause more than 50 events. We find that 116 events have the distance of background stars from the proper motion path of lens stars more than $8^{\prime\prime}$ in the reference epoch, where the maximum distance is $16.6^{\prime\prime}$, so the cone search method of expanding the search range of sources for each potential lens star can reduce the possibility of missing events.

\end{abstract}



\section{Introduction}


Gravitational lensing describes the deflection and magnification of background sources when a massive object (lens) passes in front of them.  When the lens is the stellar-mass object, the deflection is less than a few microarcseconds (mas) and the effect is referred to as microlensing. Microlensing describes the positional deflection (astrometric microlensing) and magnification (photometric microlensing) of a background source over time \citep{Paczynski,Hog1995,Miyamoto1995,Walker1995}. The photometric microlensing has been investigated by surveys such as the Optical Gravitational Lensing Experiment (OGLE) \citep{Udalski} or the Microlensing Observations in Astrophysics (MOA) \citep{Bond}, whereas the astrometric microlensing was detected for the first time only recently \citep{Sahu17,Zurlo,McGill2023}.

Astrometric microlensing provides the possibility to directly measure the mass of a single star \citep{Paczynski95,Kains2017}, it also can detect faint and compact lens such as isolated neutron stars and black holes because the luminosity of the lens is not necessarily measured. Recently, an isolated stellar-mass black hole was detected by \citet{Sahu} through astrometric microlensing based on the Hubble Space Telescope (HST) astrometry and the ground-based photometry, and for the  microlensing events \citet{Lam2022} proposed that the lens star is a compact object.

The microlensing events are intrinsically rare occurrences, e.g. the all-sky averaged value of the astrometric optical depth is $2.5 \times 10^{-5}$ \citep{Belokurov},
that depend on the alignments of source star and lens star, so predicting when and where they will occur is highly advantageous for the collection of data throughout an event.

Gaia Data has been proven to be ideal for predicting astrometric microlensing events \citep{Collaboration}, and about 25 000 sources will have a significant variation of the centroid shift during the Gaia mission period \citep{Belokurov}. Using Gaia DR1 (Gaia Data Release 1), \citet{McGill} predicted an astrometric microlensing event with the white dwarf LAWD 37 as a lens star,and recently \citet{McGill2023} measured the astrometric deflection of the background source and obtained the mass of lens star LAWD37. After Gaia DR2 (Gaia Data Release 2) was released, many research works have been done. \citet{Bramich18a} and \citet{Bramich18b} predicted the microlensing events between 25th July 2014 and the end of the century. In addition, \citet{McGill19} used the data of Gaia DR2 and Vista Variables in the Via Lactea Infrared Astrometric Catalog \citep{Smith2018} to predict two astrometric microlensing events,and \citet{McGill2019} searched for two upcoming photometric microlensing events. \citet{Mustill2018} predicted the photometric microlensing events between 2015.5 and 2035.5. \citet{Ofek2018} searched for the astrometric microlensing events between pulsars and stars in Gaia DR2. \citet{Kluter18a,Kluter18b} predicted 3914 astrometric microlensing events caused by 2875 different lenses between 2010 and 2065. In 2020, Gaia's early Data release 3 (Gaia eDR3) data was released, \citet{Kluter22} updated their prediction results and added 1758 new microlensing events between 2010 and 2066. In addition, \citet{RN22} searched for astrometric microlensing events by nearby brown dwarfs, but this work did not reveal any upcoming microlensing events. \citet{Wyrzykowski2022} identified 363 photometric microlensing events in Gaia Data Release 3 (Gaia DR3) covering the years of 2014 - 2017 in all over the sky, and \citet{Jablonska2022} found that one of the events discovered by \citet{Wyrzykowski2022} may lead to a measurable astrometric signal.

In June 2022, Gaia Data Release3 (Gaia DR3) was released, and its astrometry and broad band photometry content are the same as those of Gaia EDR3. Gaia DR3 contains 585 million sources with five-parameter astrometry (two positions, the parallax, and two proper motion components), and about 882 million sources with six-parameter astrometry, including an additional pseudocolour parameter given by \citet{Collaboration22}. Gaia DR3 also released many new data products, such as the astrophysical parameters (mass and age) of 128 million stars \citep{Creevey2022,Fouesneau2022}, and the mass estimation of some lens stars used in this paper.

In this paper, we search for microlensing events using Gaia DR3, and  potential lens stars are selected from the three types of stars, high-proper-motion stars (HPMS), nearby stars (NS) and high-mass stars (HMS). It should be noted that NS do not contain HPMS. 

In Section \ref{sec:Theoretical Background}, we briefly outline the theoretical background for astrometric microlensing. In Section \ref{sec_Data Sources}, we present the lens and background star catalogs used in the paper. In Section \ref{sec Methodology}, we detail the methods of searching for background star matched with lens star,  calculate the source-lens closest approach, and estimate lens masses. We present our results in Section \ref{sec Results}. In Section \ref{sec Summary and Conclusions}, we give summary and conclusions.\\

\section{Theoretical background}
\label{sec:Theoretical Background}

\subsection{Astrometric Microlensing Signals}
\label{sec:Basic facts} 

The theory of astrometric microlensing is described in detail in the literature \citep{Belokurov,Dominik,Paczynski96} and we briefly introduce important concepts and equations directly relevant to this paper.

When the background source (S), the lens (L), and the observer are perfectly collinear, the lensed image of the source will form a so-called Einstein ring. The characteristic size of this ring is given by the Einstein radius as
\begin{equation}
	\label{Einstein}
	\theta_{\mathrm{E}} =\sqrt{\frac{4GM_{L}}{c^2}\frac{D_{S}-D_{L}}{D_{L}D_{S}}},
\end{equation}
where $G$ is the gravitational constant, $c$ is the speed of light, $M_L$ is the mass of the lens, $D_S$ and $D_L$ are the distances between the observer and the background source or the lens.

The angular position vector $\boldsymbol{\varphi}$ on the celestial sphere for star is given:
\begin{equation}
	\label{angular_position}
	\boldsymbol{\varphi}=\left(\begin{array}{l}
		\alpha_{0} \\
		\delta_{0}
	\end{array}\right)+\left(\begin{array}{c}
		\mu_{\alpha^{*}} / \cos \delta_{0} \\
		\mu_{\delta}
	\end{array}\right) \cdot \left(t-t_{r e f}\right)+\varpi \cdot \boldsymbol{P}(t),
\end{equation}
where $\boldsymbol{P}(t)$ is expressed by
\begin{equation}
	\label{P(t)}
	\boldsymbol{P}(\mathrm{t})=\left(\begin{array}{c}
		{\left[X(t) \sin \alpha_{0}-Y(t) \cos \alpha_{0}\right] / \cos \delta_{0}} \\
		X(t) \cos \alpha_{0} \sin \delta_{0}+Y(t) \sin \alpha_{0} \sin \delta_{0}-Z(t) \cos \delta_{0}
	\end{array}\right),
\end{equation}
and $\alpha_{0}$, $\delta_{0}$,$\mu_{\alpha}$, $\mu_{\delta}$ and $\varpi$ represent the right ascension, declination, proper motion in right ascension direction, proper motion in declination direction and annual parallax respectively. $t_{ref}$ is the reference epoch, $X(t)$, $Y(t)$ and $Z(t)$ are the Cartesian barycentric Solar-system coordinates in au of the earth on the ICRF at time $t$. In this paper, the above coordinate are calculated by the astropy PYTHON package from NASA JPL's Horizons Ephemeris \citep{AstropyCollaboration, AstropyCollaboration18,AstropyCollaboration2022}.

Let $\varphi_S$ and $\varphi_L$ represent the angular positions of the source and the lens, respectively. One can define the dimensionless distance vector as
\begin{equation}
	\label{u}
	\mathbf{u}=\frac{\boldsymbol{\varphi_S}-\boldsymbol{\varphi_L}}{\theta_E}.
\end{equation}
Usually two images of the source are observed when the source is not perfectly aligned with the lens. The bright image (+) is close to the source and the faint image (-) is close to the lens. Their distance relative to the lens are given by
\begin{equation}
	\theta_{\pm}=\frac{u\pm\sqrt{(u^2+4)}}{2}\cdot\theta_E ,
\end{equation}
where $u=|\mathbf{u}|$.
When the separation of the images is too small to be resolved, only the centroid of light formed by the images can be measured. This can be described by
 \begin{equation}
	\theta_C=\frac{u^2+3}{u^2+2}u\cdot\theta_E ,
\end{equation}
and the corresponding shift is the astrometric signal given by
\begin{equation}
	\label{centroid}
	\delta\theta_C=\frac{u}{u^2+2}\cdot\theta_E .
\end{equation}
The maximum shift of the center of light occurs at $u=\sqrt{2}$.

When the lens is luminous and unresolved from the source , the shift between lensed and unlensed position (position of the combined center of light) can be determined by
\begin{equation}
	\label{centroid_lum}
	\delta\theta_{C,lum}=\frac{u\cdot\theta_E}{1+f_{LS}}\cdot\frac{1+f_{LS}(u^2+3-u\sqrt{u^2+4})}{u^2+2+f_{LS}u\sqrt{u^2+4}},
\end{equation}
where $f_{LS}$ is the flux ratio between the lens and the source.
When the separation between the lens and the source is large enough for lens and the brighter image (+) to be resolved, the brighter image (+) will be measured. The corresponding shift compared to the unlensed position can be expressed by
\begin{equation}
	\label{+}
	\delta\theta_{+}=\frac{\sqrt{u^2+4}-u}{2}\cdot\theta_{E}.
\end{equation}
Therefore $\delta\theta_{C,lum}$, $\delta\theta_C$ and $\delta\theta_+$ are the astrometric microlensing signals.

\subsection{The change of astrometric microlensing signal with time}
\label{sec	The change of astrometric microlensing signal over time}

Refer to the derivation method of \cite{Dominik}, we deduce the relationships between the diurnal variation of the astrometric signals ($\delta\theta_{C,lum}$, $\delta\theta_C$ and $\delta\theta_+$) and the parameters of $u$, $f_{LS}$ , $\mu_{LS}$ and $\varpi_{LS}$, where $\mu_{LS}$ and $\varpi_{LS}$ are the relative proper motion and parallax between the source and the lens, respectively. First we obtain the following equation from the equations (\ref{u}) and (\ref{+}):
\begin{equation}
	\begin{array}{c}
		\label{d_+_sep}
		\frac{d\delta \theta_+ }{d \varphi } =\frac{d\delta  \theta _+}{du}\cdot \frac{du}{d \varphi }
		=\frac{u}{2\cdot \sqrt{u^2+4 } } -\frac{1}{2},
	\end{array}
\end{equation}
and then we have the equations from the equations (\ref{angular_position}) and (\ref{P(t)}):
\begin{equation}
	\begin{array}{l}
	
		\label{d_sep_dt}
		\frac{d\varphi }{dt} \approx \left |\frac{d\boldsymbol{\varphi }}{dt}\right |=
		\sqrt{(\Delta \alpha _{\mu, \varpi })^2+
		(\Delta \delta _{\mu, \varpi })^2 }  ,\\
    	\Delta \alpha _{\mu, \varpi }=\mu_{\alpha_{LS}^* } + \varpi_{LS} \cdot \Delta P_{\alpha^*},\\
    	\Delta \delta _{\mu, \varpi }=\mu_{\delta_{LS}  } + \varpi_{LS} \cdot \Delta P_{\delta }

	\end{array}
\end{equation}
where $\Delta P_{\alpha ^*}=\Delta x\cdot sin(\alpha_{0,L} )-\Delta y\cdot cos(\alpha_{0,L} )$ , $\Delta P_{\delta }=\Delta x\cdot cos(\alpha_{0,L})\cdot sin(\delta_{0,L} )+\Delta y\cdot
sin(\alpha_{0,L})\cdot sin(\delta_{0,L})-\Delta z\cdot cos(\delta_{0,L})$, and $\Delta x=-0.0171 AU $, $\Delta y=-0.0036 AU$,  $\Delta z=-0.0016 AU$ are the daily variation of x, y, z respectively. We take the daily variation from 2455199.5JD to 2455200.5JD near the perihelion, and set $\mu_{\alpha_{LS}^*}=\mu_{\alpha^*_L}-\mu_{\alpha^*_S}$, $\mu_{\delta_{LS} }=\mu_{\delta_L}-\mu_{\delta_S}$, and $\varpi_{LS} =\varpi_L-\varpi _S$. It should be noted that since the projection on the parallactic motion is position depended, the equation underestimates the effect for stars with $\alpha \sim 0^\circ$ or $\alpha \sim 180^\circ$. However, the effect only causes a small change in $\frac{d\varphi }{dt}$, so it does not affect our conclusion. Then we obtain
\begin{equation}
	\begin{array}{c}
		\label{+_dt}
		\frac{d\delta \theta_+ }{dt}=\frac{d\delta \theta_+ }{d \varphi } \cdot \frac{d\varphi }{dt}=\left (\frac{u}{2\cdot \sqrt{u^2+4 } } -\frac{1}{2} \right ) \cdot \frac{d\varphi }{dt} .
	\end{array}
\end{equation}

According to the equations (\ref{centroid}), we then have
\begin{equation}
	\label{C_t}
	\frac{d \delta \theta_{C}}{d t}=\frac{d\delta \theta_C }{d \varphi } \cdot \frac{d\varphi }{dt}= \frac{2-u^{2}}{\left(u^{2}+2\right)^{2}} \cdot \frac{d\varphi }{dt},
\end{equation}

According to the equations (\ref{+_dt}),(\ref{C_t})), we obtain that the diurnal variation of the astrometric signal depending on $\mu_{LS}$ and $\varpi_{LS}$  varies with $u$. For $\delta\theta_{C}$ and $\delta\theta_{+}$ , their maximum diurnal variations increase with $\mu_{LS}$ and $\varpi_{LS}$. The maximum diurnal variation of  $\delta\theta_{+}$ can exceed 0.1mas only when $u<5$. Therefore, for most astrometric microlensing events, the change of their astrometric signals within 24 hours (or even longer) cannot be detected by space telescope, e.g. Hubble Space Telescope \citep{HST} and James Webb Space Telescope \citep{JWST}, we do not need high accuracy in finding the closest approach ($u_0$ and $t_0$) because the shift doesn't vary by an appreciable amount on a small timescale (e.g. a daily timescale). We just need to encrypt the calculation points appropriately during certain periods.

In this paper, according to the characteristic of that the astrometric signals of microlensing events change slowly with time, we only take a small number of data points when calculating the stellar trajectory, and can find the closest approach ($u_0$ and $t_0$). More detailed description is given in Section \ref{sec:the source-lens closest approach}.

\section{Data Sources}
\label{sec_Data Sources}

\subsection{Lens star selection}
\label{sec	Lens star selection}
High proper motion stars are typically selected as potential lenses \citep{Kluter18b,Kluter22,McGill19} because they traverse large areas of the sky in a given time period which increases the rate of close alignments to background sources. In addition, according to equation (\ref{Einstein}), for the events  with $D_S \gg D_L$, the closer the lens star is to the Earth or the larger its mass is, the larger the radius of the Einstein ring is. For a given lens-source angular separation, the larger $\theta_E$ corresponds to a larger astrometric signal, the larger the observed signal is, the higher the probability observed by the telescope is. Therefore, in this paper, the potential lens stars are selected from three types of stars, HPMS, NS and HMS (as shown in Table \ref{conditions_lens}).

The potential lens star comes from Gaia DR3 and must have position, parallax, and proper motion. To ensure a good astrometric solution, we require the parallaxes with relatively small errors parallax\_over\_error>5. However, the source with parallax\_over\_error>5 may also contain about 1.6\% spurious astrometric solutions \citep{Fabricius2021}. In order to classify valid and spurious astrometric solutions, \citet{Rybizki22} train a neural network to get a parameter of "astrometric fidelity"  taken between 0 and 1 for 1.47 billion sources (e.g. five-parameter solutions or six- parameter solutions) in Gaia eDR3. A value of 1.0 means a perfectly trustable solution, and the lowest value of 0.0 indicates a lot of issues in the astrometric solution. Using this parameter to eliminate spurious astrometric solutions is more effective than simple cutting. \citet{Rybizki22} suggested "In most regimes, the use of the astrometric fidelitiy should yield a purer and more complete sample of sources with reliable astrometric solutions." Because the astrometric data of Gaia DR3 is the same as that of Gaia eDR3, the "astrometric fidelity" parameter is still applicable in Gaia DR3. In this paper, the lens with the "astrometric fidelity" parameter greater than 0.8 (fidelity\_v2>0.8) are selected. This limit of the parameter is the condition that all potential lens stars need to meet.

For HPMS, $\mu\ge 100mas/yr$,  where $\mu$ is the total proper motion, their number is about 470000.

For NS,  $\varpi>10mas$, and $\mu < 100mas/yr$ is also needed to avoid the overlap with HPMS. Their number is about 160000.

For HMS, their mass provided by Gaia DR3 is limited as  $M_L>5M_{\odot}$, and $\mu<100mas/yr$ and $\varpi<10mas$ are needed to avoid the overlap with HPMS and NS. Their number is about 190000.

\begin{table}
	\centering
	\caption{The conditions of the potential lens stars}
	\label{conditions_lens}
	\begin{tabular}{lll}
		\hline
		\hline
		\multicolumn{1}{l|}{HPMS}          & \multicolumn{1}{l|}{NS}                      & HMS                     \\ \hline
		\multicolumn{1}{l|}{$\mu\geq$100mas/yr} & \multicolumn{1}{l|}{$\varpi$\textgreater{}10mas}    & $\varpi$\textless{}10mas      \\
		\multicolumn{1}{l|}{}              & \multicolumn{1}{l|}{$\mu$\textless{}100mas/yr} & $\mu$\textless{}100mas/yr \\
		\multicolumn{1}{l|}{}              & \multicolumn{1}{l|}{}                        & $M_{L}>5 M_{\odot}$ \\
		\multicolumn{1}{l|}{$\sim$470000}  & \multicolumn{1}{l|}{$\sim$160000}            & $\sim$190000            \\ \hline
		\multicolumn{3}{l}{parallax\_over\_error\textgreater{}5}                                                    \\
		\multicolumn{3}{l}{phot\_g\_mean\_mag IS NOT NULL}                                                          \\
		\multicolumn{3}{l}{astrometric\_params\_solved=31(OR astrometric\_params\_solved=95)}                       \\
		\multicolumn{3}{l}{fidelity\_v2\textgreater{}0.8}                                                                \\ \hline
		
		\multicolumn{3}{l}{Notes. $\mu$ is the total proper motion, $\varpi$ is parallax,}\\
		\multicolumn{3}{l}{$M_L$ is the mass estimation of star (mass\_flame) in the appendix }\\
		\multicolumn{3}{l}{Astrophysical parameters provided by Gaia DR3,  }\\
		\multicolumn{3}{l}{parallax\_over\_error is parallax divided by its standard deviation, }\\
		\multicolumn{3}{l}{ phot\_g\_mean\_mag is G-band mean magnitude,}\\
		\multicolumn{3}{l}{astrometric\_params\_solved is astrometric solutions, }\\
		\multicolumn{3}{l}{astrometric\_params\_solved=31 is five-parameter solutions,}\\
		\multicolumn{3}{l}{astrometric\_params\_solved=95 is six-parameter solutions,}\\
		\multicolumn{3}{l}{fidelity\_v2 is "astrometric fidelity" parameter from \citet{Rybizki22} }\\
	\end{tabular}
\end{table}

\subsection{Background star selection}
\label{sec	Background stars star selection}

Background star (BGS) data are also taken from Gaia DR3. The selection conditions are listed in Table \ref{conditions_background}. The condition is fidelity\_v2>0.7 for background  stars with five - and six-parameter solutions. For background  stars with two-parameter solution, \citet{Rybizki22} did not provide "astrometric fidelity" parameter, and Gaia DR3 did also not provide the ruwe parameter \citep{Lindegren2021}. However, we can estimate the ruwe parameter according to the Gaia DR3 documentation (see the section 20.1.1 of Gaia DR3 \footnote{\url{https://gea.esac.esa.int/archive/documentation/GDR3/Gaia_archive/chap_datamodel/sec_dm_main_source_catalogue/ssec_dm_gaia_source.html}}) , where the equations include
\begin{equation}
	\begin{array}{c}
		\label{astrometric_gol_al}
		astrometric\_gol\_al=(9 \cdot v / 2)^{1 / 2} \cdot \left[ruwe^{2 / 3}+2 /(9 \cdot v)-1\right],
	\end{array}
\end{equation}
and
\begin{equation}
	\begin{array}{c}
		
		v=astrometric\_n\_good\_obs\_al-N.
	\end{array}
\end{equation}
Then the ruwe parameter is given by
\begin{equation}
	\begin{array}{c}
		\label{ruwe}
		 ruwe =\left(\frac{astrometric\_gof\_al }{\sqrt{9 \cdot v / 2}}-\frac{2}{9 \cdot v}+1\right)^{3 / 2},
	\end{array}
\end{equation}
where $astrometric\_gof\_al$ is the goodness of fitting statistic of model along-scan observations, $v$ is the number of degrees of freedom for a source update, and $N=5$. Therefore, the condition for background stars with two-parameter solution is $ruwe<2$ and $\sqrt{\sigma_{\alpha^*}^2+\sigma_\delta^2}<10mas$. In the subsequent calculations, we set the parallax and proper to be 0 for background stars with two-parameter solution and assumed standard errors of $\sigma_{\alpha^*_{S},\delta_{S}}=5 mas/yr$ and $\sigma_{\varpi_S}=2 mas$. Roughly 90\% BGSs with five or six parameter solution have the standard errors in proper motion and parallaxe below the above values. 

\begin{table}
	\centering
	\caption{The conditions of background stars}
	\label{conditions_background}
	\begin{tabular}{cc}
		\hline
		\hline
		two-parameter solution & five - and six-parameter solutions \\
		\hline
		\begin{tabular}[c]{@{}c@{}} ruwe<2,\\ $\sqrt{\sigma_{\alpha^*}^2+\sigma_\delta^2}<10mas$ \end{tabular}
		 & fidelity\_v2\textgreater{}0.7    \\
		\hline
	\end{tabular}
\end{table}

\section{Predicting microlensing events}
\label{sec Methodology}

\begin{figure*}
	\includegraphics[width=2.0\columnwidth]{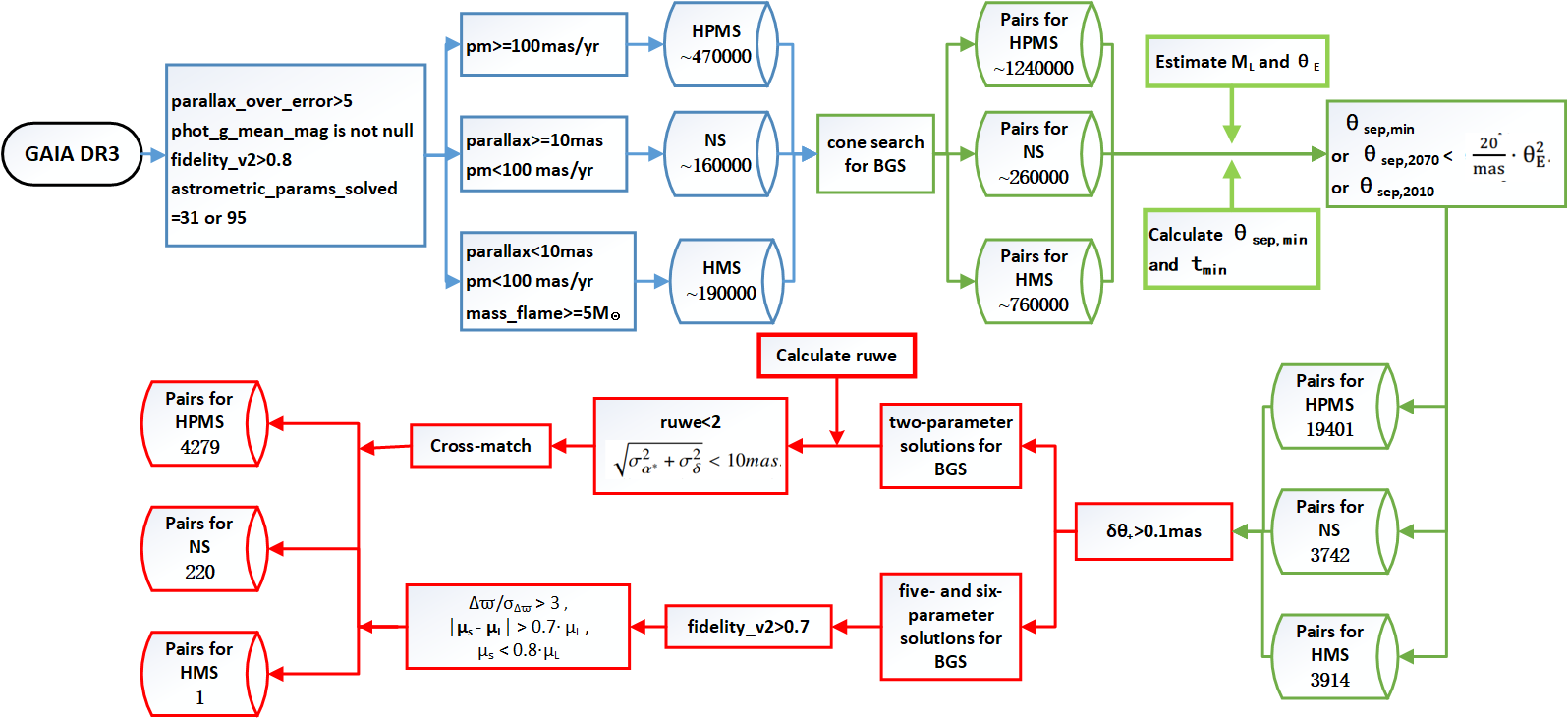}
	\caption{Illustration of the selection processes. They are the selection of potential lens stars (blue), the selection of background stars (green), the determination of the closest approach, and the exclusion of co-moving stars and the estimation of the expected microlensing effect (red).}
	\label{Fig_selection process}
\end{figure*}

\begin{figure*}
	\includegraphics[width=2.0\columnwidth]{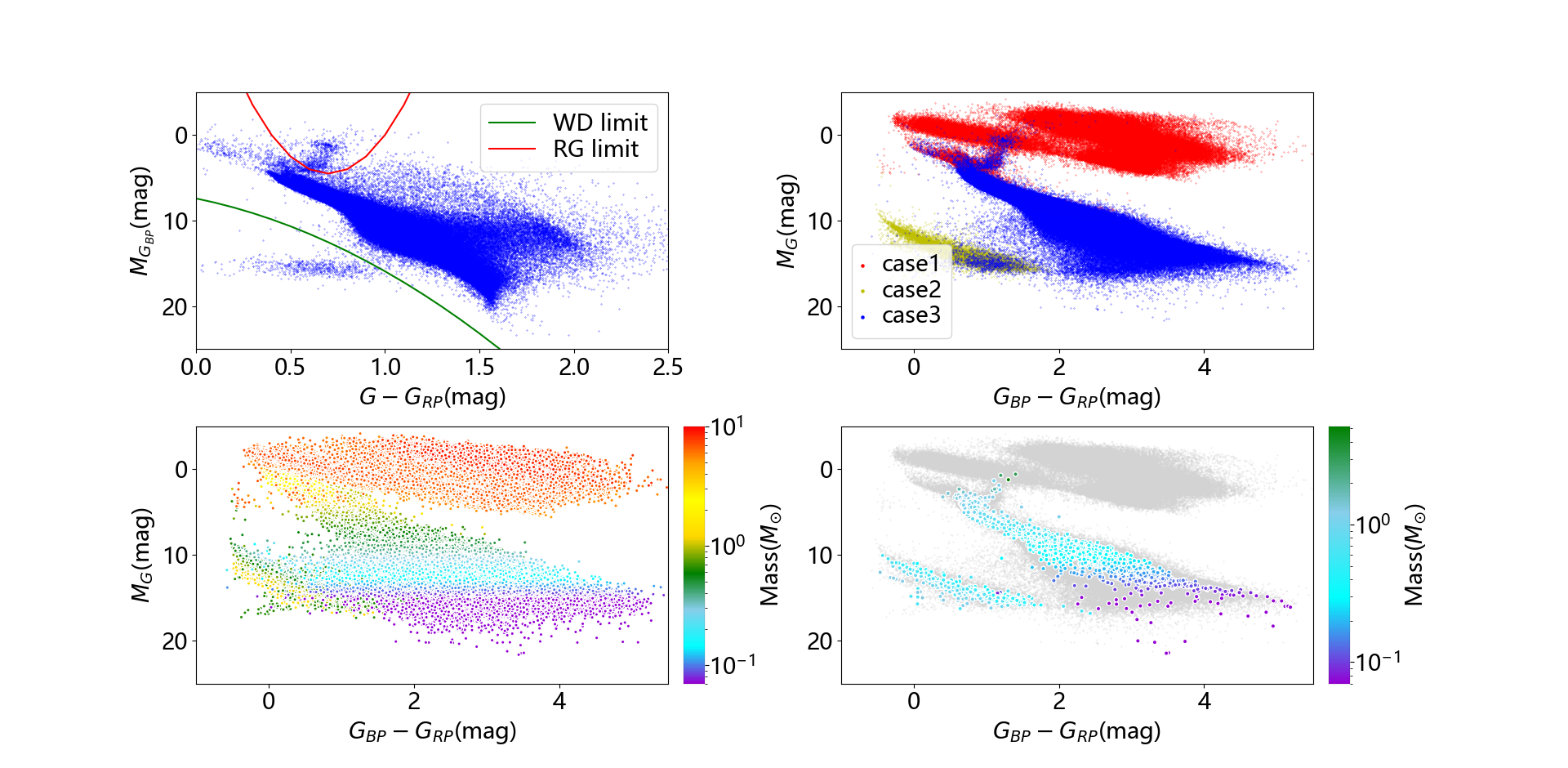}
	\caption{Color-magnitude Diagrams. Upper-left panel: Color-magnitude diagram of the third case potential lens with $G, G_{BP}, G_{RP}$ from Section \ref{sec:Mass estimation of lens stars}, $M_{G_{BP}}$ is the absolute magnitude at $G_{BP}$ band, the stars below the green line are considered to be WDs, and the stars above the red line are considered to be RGs. Upper right panel: Hertzsprung-Russell diagram of $M_{G}$ versus $G_{BP}-G_{RP}$ for all potential lens with $G, G_{BP}, G_{RP}$ from Section \ref{sec:Mass estimation of lens stars}. The red dots indicate the first case, the yellow indicate the second case and the blue indicate the third case. Lower left panel: Same as the upper right panel for all potential lens with $G, G_{BP}, G_{RP}$. The lens masses are indicated by the colour of the points (see the scale at the right of the panel). Lower right panel: Same as lower left panel, the grey indicate all potential lens with $G, G_{BP}, G_{RP}$ and the colour dots indicate the lenses of predicted events.}
	\label{Fig_lens_RH}
\end{figure*}

The selection process is shown in Figure \ref{Fig_selection process}. First, the potential lens (blue) are selected, then the background stars are matched (green), and finally the predicted events that meet the conditions are selected (red).

\subsection{Initial lens-source matching}
\label{sec Initial lens-source matching}

For $\sim 820000$ potential lens stars, we query for all sources using the cone search method, where the search time range is J2010.0 $\sim$ J2070.0. The center of the circle is the position of the lens at the reference epoch (J2016.0) of Gaia DR3. For HPMS, the search radius is $T\cdot\mu+6^{\prime\prime}$, and $\sim 1240000$ star pairs are found. For NS and HMS, the search radius is $T\cdot\mu+8^{\prime\prime}$, and $\sim 260000$ and $\sim 760000$ star pairs are found respectively. $ T=54 yr$ is from J2070.0 to J2016.0. To avoid missing events as much as possible, we use the cone search method that has larger searching area than that of the rectangular search method \citep{Kluter18b,Kluter22,McGill19}, and the subsequent steps can further constrain the star pairs.

Now the number of founded pairs are still large, we further select the pair of stars according to the characteristics of their relative motion. The relative motion of the lens-background pair can be approximated as linear motion (without considering parallax), and the relative position ($\boldsymbol{\Delta \varphi}$) of the star pair with time can be given by the equation (\ref{angular_position}) as
\begin{equation}
	\begin{array}{c}
		\label{Relative angular distance}
		\boldsymbol{\Delta \varphi} \approx\left(\begin{array}{c}
			\alpha_{0 L}-\alpha_{0 S} \\
			\delta_{0 L}-\delta_{0 S}
		\end{array}\right)+\left[t-t_{r e f}\right]\left(\begin{array}{c}
			{\left[\mu_{\alpha^{*} L}-\mu_{\alpha^{*} S}\right] / \cos \delta_{0 S}} \\
			\mu_{\delta L}-\mu_{\delta S}
		\end{array}\right) \\
		=\left(\begin{array}{c}
			\alpha_{0, \text { LS }} \\
			\delta_{0, \text { LS }}
		\end{array}\right)+\left[t-t_{\text {ref }}\right]\left(\begin{array}{c}
			\mu_{\alpha^{*}, \text { LS }} / \cos \delta_{0 S} \\
			\mu_{\delta, \text { LS }}
		\end{array}\right),
	\end{array}
\end{equation}
where $\alpha_{0L}$, $\delta_{0L}$, $\alpha_{0S}$ and $\delta_{0S}$ are the right ascensions and declinations of the lens star and the background star at the reference epoch, respectively. $\mu_{\alpha^\ast L}$, $\mu_{\delta L}$, $\mu_{\delta S}$  and $\mu_{\alpha^\ast S}$ are the proper motions in right ascension and declination directions of the lens star and the background star. $\alpha_{0,LS}$ and $\delta_{0,LS}$ are the relative right ascension and relative declination of the star pair, $\mu_{\alpha^\ast,LS}$ and $\mu_{\delta,LS}$ are proper motions in right ascension and declination directions of the star pair.

Therefore, the relative angular distance ($\theta_{sep}$) of the star pair can be approximated as
\begin{equation}
	\begin{array}{c}
		\begin{aligned}
			
			\label{sep}
			\theta_{sep}\approx \sqrt{\left ( \Delta \alpha \cdot cos\delta _{0S} \right ) ^{2}+\left ( \Delta \delta  \right )^{2}  },
						
		\end{aligned}
	\end{array}
\end{equation}
where $\Delta \alpha = \alpha_{0,LS} +\left(t-t_{r e f}\right) \cdot \mu_{\alpha^{*}, LS}/cos\delta _{0S}$ and $\Delta \delta=\delta_{0, LS}+\left(t-t_{r e f}\right) \cdot \mu_{\delta, LS}$. Based on $ \frac{\partial \theta_{sep}}{\partial t} =0$, we can get the minimum value $\theta_{sep,min}$ of $\theta_{sep}$ and its corresponding time $t_{min}$ as
\begin{equation}
	\begin{array}{c}
		\label{sep_minimum}
		\theta_{sep,min} = \left |\frac{\mu_{\delta,LS}\cdot \alpha_{0,LS} \cdot \cos \delta_{0 S} -\mu_{\alpha^{*},LS} \cdot \delta_{0,LS}}{\sqrt{\mu_{\alpha^{*},LS}^{2}+\mu_{\delta,LS} ^{2}}}\right | ,
	\end{array}
\end{equation}
and
\begin{equation}
	\begin{array}{c}
		\label{t_minimum}
		t_{min} = -\frac{\alpha_{0, LS} \cdot \cos \delta_{0 S} \cdot \mu_{\alpha^{*},LS}+ \delta_{0,LS}\cdot \mu_{\delta,LS}}{\mu_{\alpha^{*},LS}^{2}+\mu_{\delta,LS} ^{2}}+t_{ref}.
	\end{array}
\end{equation}

For all star pairs searched above, we use the equation (\ref{t_minimum}) to calculate $t_{min}$. If $t_{min}$ is within the range of J2010.0-J2070.0, the equation (\ref{sep_minimum}) is used to calculate $\theta_{sep,min}$. If $t_{min}>2070$,  the relative angular distance $\theta_{sep,2070}$ is calculated at $t=2070$ according to the equation (\ref{sep}). If $t_{min}<2010$, the relative angular distance $\theta_{sep,2010}$ is calculated at $t=2010$ based on the equation (\ref{sep}). If $\theta_{sep,min}$, $\theta_{sep,2070}$ or $\theta_{sep,2010}$ is less than $\frac{20}{mas} \cdot \theta ^2_E$, e.g. $\delta\theta_{+,mas}>0.05mas$ ($\delta \theta_+\sim \frac{\theta_E}{u}=\frac{\theta_{E^2}}{\varphi}$), where $\delta\theta_{+,max}$ is the approximate value of the maximum value of $\delta\theta_+$ in the range J2010.0-J2070.0, the star pair can be initially matched, otherwise the star pair is excluded. $\theta_E$ is calculated according to the equation (\ref{Einstein}), where the lens star masses are estimated in Section \ref{sec:Mass estimation of lens stars} .

After the application the above criteria, we find 19401, 3742 and 3914 pairs for HPMS, NS and HMS respectively. It is noted that there is no restriction on the background star at this time.

\subsection{Mass estimation of lens stars}
\label{sec:Mass estimation of lens stars}

The masses of these potential lens stars are estimated in three cases :

First case, the masses are searched in the "Astrophysical parameters" table of 128 million stars released by Gaia DR3 \citep{Creevey2022,Fouesneau2022}.  The mass estimation of all HMS comes from this table. In addition, the masses of about 40000 HPMS and NS are given in this table (see upper-right panel of Figure \ref{Fig_lens_RH}, the mass estimation of the red dots in the panel comes from this table).

Second case, the masses are estimated by matching the white dwarf catalog \citep{Gentile2021}. The catalog provides the parameter 'PWD' as a measure of the probability of the source being a white dwarf. In this paper, the potential lens stars that do not belong to the first case are matched with the white dwarf star catalog. About 13000 potential lens stars are matched with the catalog (see upper-right panel of Figure \ref{Fig_lens_RH}, the mass estimation of the yellow dots in the panel comes from the white dwarf star catalog), where  99\% potential lens stars with 'PWD'>0.7 . \citet{Gentile2021} listed three types of estimated mass: M\_H, M\_He and M\_mix corresponding to the mass estimated by pure-H, pure-He, mixed hydrogen/helium (H/He) compositions model, respectively. The order of use for mass estimation is M\_H, M\_He and M\_mix. If there is no mass estimation in the first order,  the second order is used, and so on. The lens ratios using  M\_H, M\_He and M\_mix are $\sim 96\%$, $\sim 1\%$, and $\sim 2\%$, respectively. According to the statistics of lens, $\sim 93\%$ of lens stars have a difference between M\_H and M\_He less than $3\sigma$, and  $\sim 70\%$ of lens stars have a difference between M\_H and M\_mix less than $3\sigma$. It is noted that about 200 potential lens stars have not mass estimations, we then use the classical mass of white dwarf stars as their masses, $M_{WD}=(0.65\pm 0.15) M_\odot $.

Third case, the masses are estimated by the mass-luminosity relations. For potential lens stars (about 270000) that do not belong to the above two cases, we use the method of \citet{Kluter22} to estimate the mass. These lens stars are divided into white dwarf, red giant, brown dwarf and main sequence stars (see upper-left panel of Figure \ref{Fig_lens_RH}). The masses of white dwarfs, red giants and brown dwarfs  are estimated as $M_{WD}=(0.65\pm 0.15) M_\odot$, $M_{RG}=(1.0\pm 0.5) M_\odot $ and $M_{BD}=(0.07\pm 0.03) M_\odot $ respectively. The masses of the main sequence stars are estimated using the mass-luminosity relations, where a 10 per cent error is assumed.

It should be noted that mass estimation is important as it can affect follow up planning decisions \citep{McGill2019}. If we can find a more accurate mass estimate from the literature, we prefer to choose it for improving the accuracy of the prediction.

\subsection{the source-lens closest approach}
\label{sec:the source-lens closest approach}
\begin{figure}
	\includegraphics[width=\columnwidth]{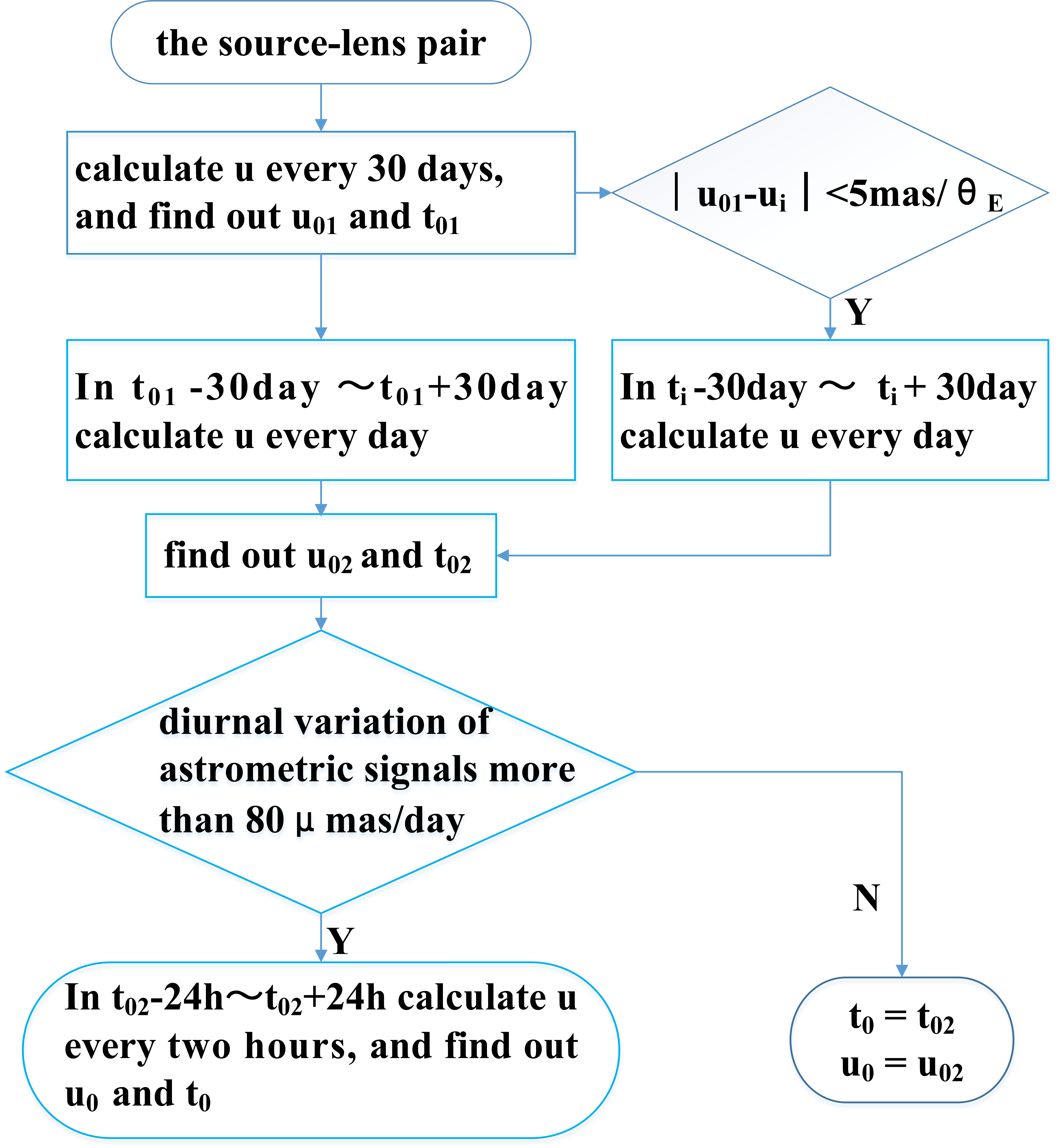}
	\caption{Flow chart of calculating the source-lens closest approach.}
	\label{Fig_flow}
\end{figure}

In this paper, the source-lens closest approach are searched by calculating the relative motion of the star pair. According to Section \ref{sec	The change of astrometric microlensing signal over time}, the signal of astrometric microlensing events changes slowly with time. Therefore, we only take a small number of data points when calculating the trajectory of the star, and only encrypted sampling is carried out in a short period of time when the signals change rapidly. The specific steps are the following as (see Figure \ref{Fig_flow}):

Step 1: we use the equations (\ref{angular_position}) and (\ref{P(t)}) to calculate the source-lens separation ($u$) every 30 days, where there are 732 data points in the range of J2010.0-J2070.0. From these data points we find the minimum source-lens separation ( $u=u_{01}$) and the time of closest approach ( $t=t_{01}$) between the source and the lens. If the difference between the source-lens separation ($u_{i}$) and the minimum source-lens separation $u_{01}$ is less than $\frac{5mas}{\theta_{E}}$, we record its corresponding time as $t_{i}$, where i represents the ith data point.

Step2:  we calculate the source-lens separation every day from 30 days before $t_{01}$ to 30 days after $t_{01}$, and from 30 days before $t_{i}$ to 30 days after $t_{i}$ , where there are 60 $\sim$ 120 data points. From these data points we find the minimum source-lens separation ($u=u_{02}$) and the time of closest approach ($t=t_{02}$) between the source and the lens. According to the equations (\ref{C_t}) and (\ref{+_dt}), we calculate  $\frac{d \delta \theta_{C}}{d t}$, and $\frac{d \delta \theta_{+}}{d t}$ with $u=u_{02}$. If the absolute value of any one of them is greater than $80\mu as/day$, we go to step 3. Otherwise, the calculation ends, and we obtain the results of $u_0=u_{02}$ and $t_0=t_{02}$.

Step3: we calculate the source-lens separation every two hours from 24 hours before $t=t_{02}$ to 24 hours after $t=t_{02}$, where there are 24 data points. From these data points we find the minimum source-lens separation ($u=u_{0}$) and the time of closest approach ($t=t_{0}$) between the source and the lens.

\subsection{Checking for microlensing events}
\label{sec:Checking for microlensing events}

\begin{figure}
	\includegraphics[width=\columnwidth]{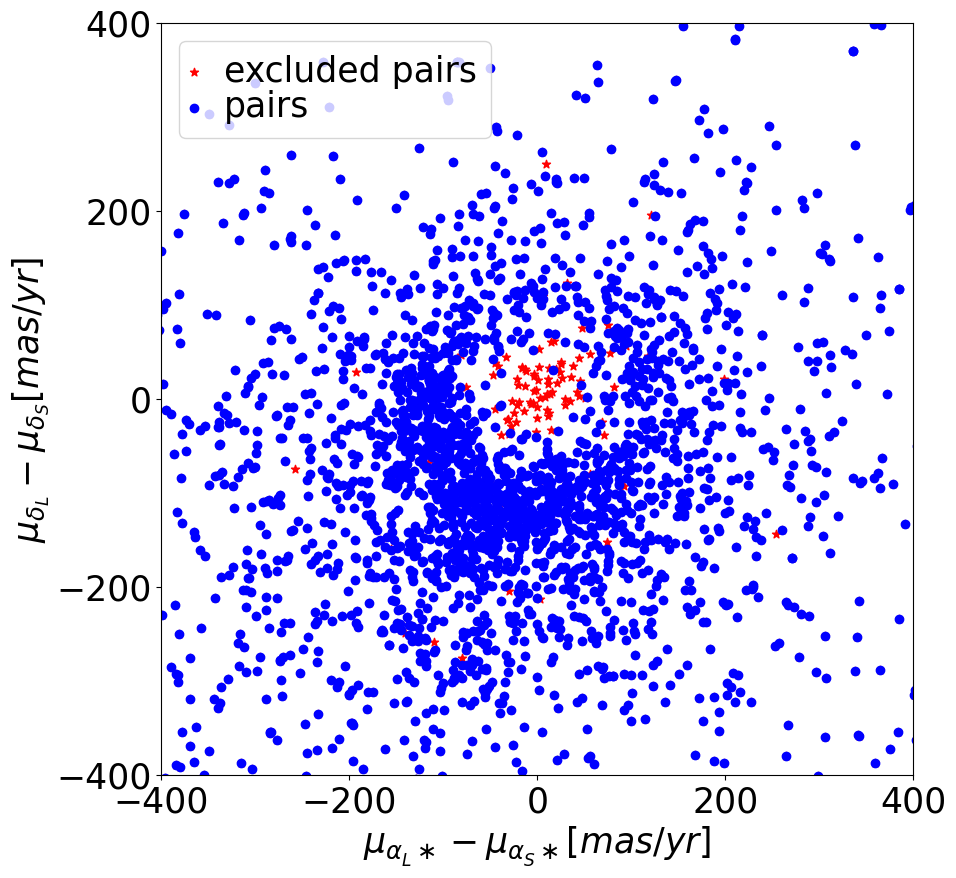}
	\caption{The proper motion difference between the lens star and the background star. The red points are star pairs that do not meet the conditions (\ref{d_parallax}) and (\ref{cut pm}). The blue dots are the final selected star pairs.}
	\label{Fig_exclude stars}
\end{figure}

Using $u_0$ and $t_0$ obtained in Section \ref{sec:the source-lens closest approach}, we calculate the observation signal $\delta\theta_+$ through the equations (\ref{Einstein}), (\ref{u}) and (\ref{+}), and find out the events with $\delta\theta_+>0.1mas$ and the background star meeting the conditions in Table \ref{conditions_background}. For these events, if the background star has a five - or six-parameter solution, it is necessary to exclude binary stars or co-moving stars. We refer to the conditions of excluding these star pairs used by many authors \citep{Bramich18a,Bramich18b,Kluter22} and the method of selecting binary stars by \citet{El-Badry2021}, and retain the star pairs meeting the following conditions (as shown in Figure \ref{Fig_exclude stars}):
\begin{equation}
	\begin{array}{c}
		\label{d_parallax}
		\frac{\Delta \varpi}{\sigma_{\Delta \varpi}}=\frac{\varpi_{L}-\varpi_{S}}{\left(\sigma_{\varpi_{L}}^{2}+\sigma_{\varpi_{S}}^{2}\right)^{1 / 2}}>3,
	\end{array}
\end{equation}

\begin{equation}
	\begin{array}{c}
	\label{cut pm}
	|\boldsymbol{\mu_{S}-\mu_L}| >0.7\cdot \mu_L,\\ \mu_{S}<0.8\cdot \mu_L.
    \end{array}
\end{equation}

 In addition, \citet{McGill2020} indicates that events with bright background sources($G<18 mag$) which are predicted to happen during the Gaia mission are likely not genuine.
For the background star with a two-parameter solution, we check these events using a method similar to that of \citet{Kluter22}. First, we search for matched sources by using the gaiadr3.dr2\_neighbourhood catalog \citep{Torra2021}. For one-to-many match case, we only consider the DR3-DR2 pair with the smallest angular distance. Then we only select the match that meets the following criteria as
	\begin{equation}
		\begin{array}{c}
			\Delta \varphi_{match}<400 mas,\\
			\Delta G_{match}<1 mag,
		\end{array}
	\end{equation}
where $\Delta \varphi_{match}$ and $\Delta G_{match}$ are the position and magnitude of source in DR3-DR2, respectively. We notice that 40 source have different source\_id in DR3-DR2 release data. To ensure a correct match, we check these sources again. Specifically, we match their dr2\_ source\_id to dr3\_source\_id and find that one of them is an error match. We have obtained 1281 sources with good match and 1117 sources with bad or no match. Second, we estimate the proper motion ($\mu_S$) of background stars with good match as
	\begin{equation}
		\begin{array}{l}
			\mu_{S}=\left[\left(\begin{array}{c}
				\alpha_{S, D R 3} \\
				\delta_{S, D R 3}
			\end{array}\right)-\left(\begin{array}{c}
				\alpha_{S, D R 2} \\
				\delta_{S, D R 2}
			\end{array}\right)\right] \cdot \begin{bmatrix}
			\cos \delta_{S, D R 3}
			\\
			1
		\end{bmatrix}
		 / \Delta t,
		\end{array}
	\end{equation}
where $\alpha_{S, DR3}$, $\delta_{S, DR3}$, $\alpha_{S, DR2}$ and $\delta_{S, DR2}$ are the positions of the BGS in Gaia DR3 and Gaia DR2 at reference epoch, respectively, $\Delta t$ is the difference between the catalog epochs of Gaia eDR3 and Gaia DR2 taken as 0.5 yr. We determine whether these events satisfy the equations (\ref{cut pm}).

We then find 1070 events satisfying the above conditions. Third, we still need to match the external catalogues for the events (42 events) with two-parameter solution BGS that are brighter than 18 mag and satisfy the equation (\ref{cut pm}). It is also required that the angular distance between the source of Gaia DR3 and the external catalogues is less than 1 arcsecond, and the lens and background star can match the same catalogues. Gaia DR3 catalogue includes pre-computed cross-matches with optical/near infrared photometric and spectroscopic surveys \citep{Marrese2017,Marrese2019}. These external catalogues matched with Gaia DR3 are Pan-STARRS1 DR1 \citep{Pan-STARRS1}, SkyMapper DR2\citep{SkyMapperDR2}, SDSS DR 13\citep{SDSSDR13}, URAT1\citep{URAT1}, Tycho2\citep{Tycho2}, Hipparcos-2\citep{Hipparcos-2}, 2MASS\citep{2MASS}, AllWISE\citep{AllWISE}, APASS
DR9\citep{APASSDR9}, GSC 2.3\citep{GSC2.3}, RAVE DR5\citep{RAVEDR5} and RAVE DR6\citep{RAVEDR6}. We exclude 33 events that cannot match any of the above external catalogues. Fourth, we exclude 993 events that cannot match any of the above external catalogues from 1117 events of that their background stars do not match with Gaia DR3.
	
In addition, we recalculate 360 events occurred before 2010 or after 2070 to estimate the exact epoch, and remove 75 events occurred before 2005 or after 2075.

For the final star pair, we determine the uncertainties of these predictions using a Monte Carlo method, where we draw 1000 samples from appropriate gaussian distribution for the lens with position, proper motion and parallax. We do not include any co-variances between different input parameters. It should be noted that for 95 events with $u_0<10$ and $\sigma_{u_0}/u_0>5$, we provide the lower confidence level (16\%) of $\delta\theta_+$, $\delta\theta_C$, and $\delta\theta_{C,lum}$ respectively, rather than the standard deviation.

\section{Results}
\label{sec Results}

\subsection{Predicted Astrometric-microlensing Events}
\label{sec Predicted Astrometric-microlensing Events}

\begin{figure}
	\includegraphics[width=\columnwidth]{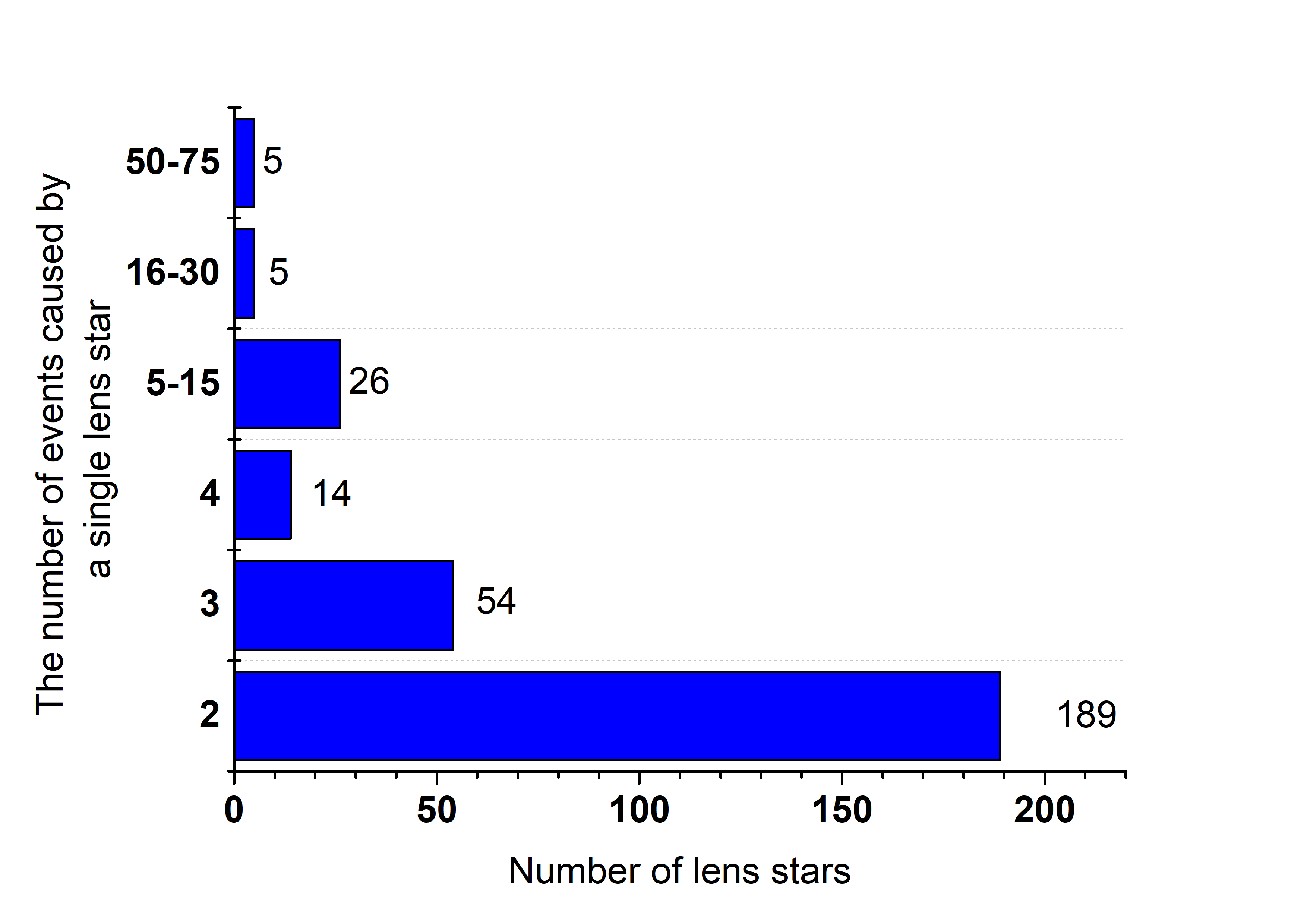}
	\caption{The statistical diagram of multiple events caused by a single lens star. The abscissa is the number of lens stars, and the ordinate is the number of events caused by a single lens star. It is noted that there are five lens stars to cause more than 50 events.}
	\label{Fig_multiple events}
\end{figure}

\begin{figure}
	\includegraphics[width=\columnwidth]{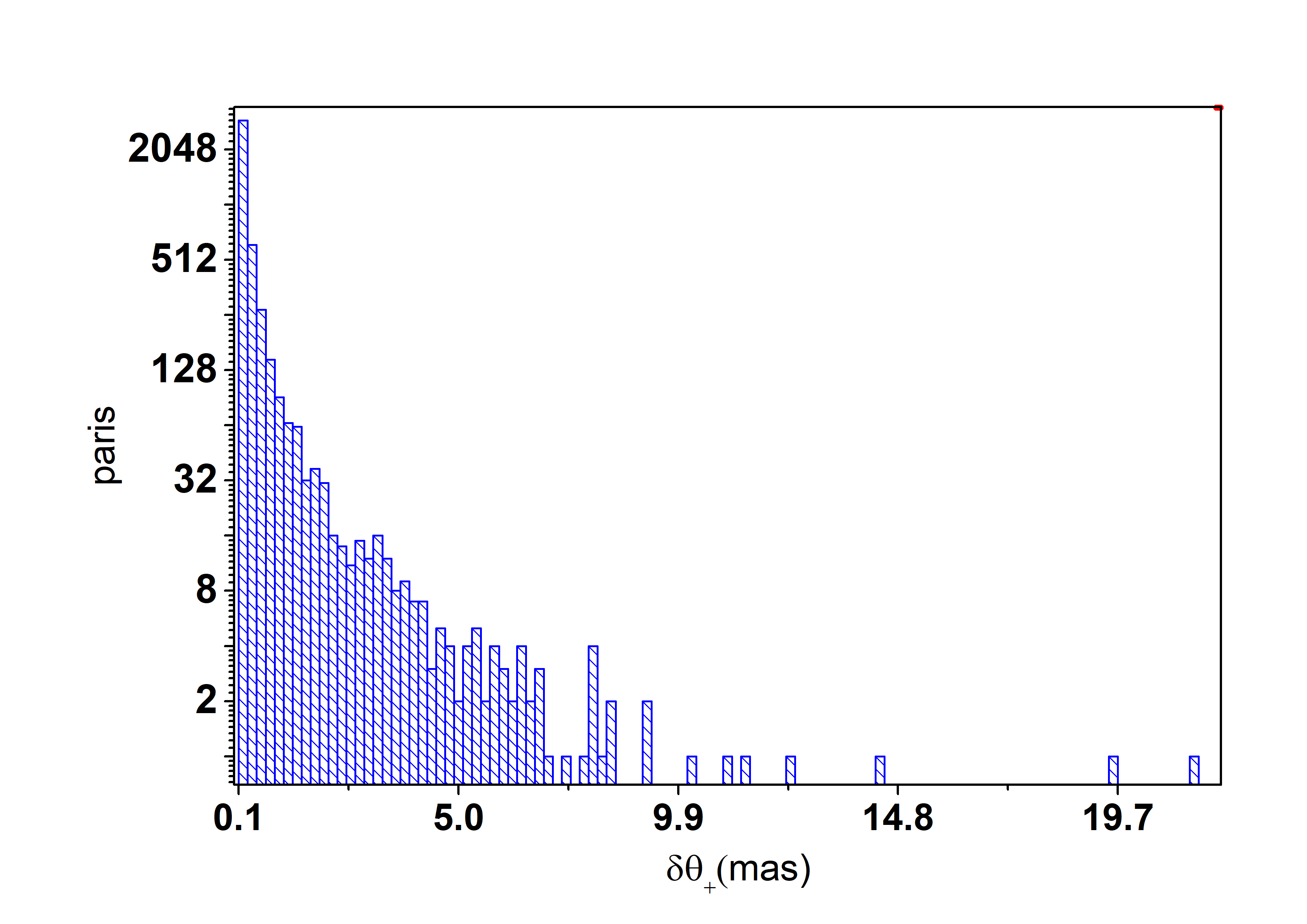}
	\caption{The histogram of $\delta\theta_{+}$  for all events. 457 events have $\delta\theta_{+}>1mas$, and $48 \%$ events have $0.1mas<\delta\theta_{+}<0.2mas$ .}
	\label{Fig_all_d+}
\end{figure}

\begin{figure}
	\includegraphics[width=\columnwidth]{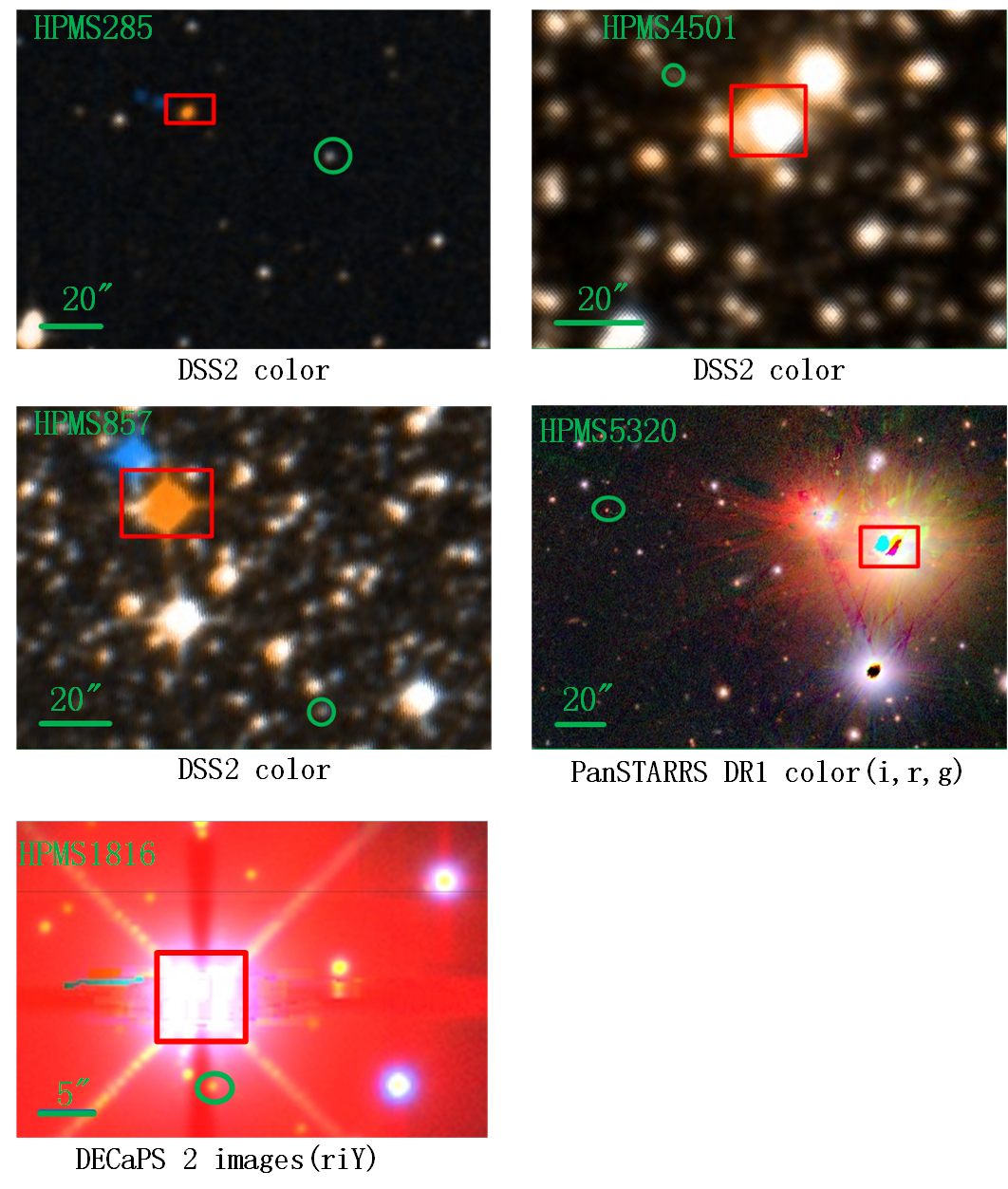}
	\caption{Cutouts for the events with $\delta\theta_{+} > 10 mas$ obtained using online observation tool: the \href{https://sky.esa.int/}{ESAsky} \citep{ESASky2017,ESASky2018} ($\sim$ J2012 for PanSTARRS DR1 color(i,r,g) and $\sim$ J1990 for DSS2 color) and the \href{http://legacysurvey.org/viewer}{Legacy Surveys}\citep{LegacySurveys} ($\sim$ J2017 for DECaPS 2 images(riY)). The lens is shown in red rectangular, and the source is indicated in green circles. There are 7 events  with $\delta\theta_{+} > 10 mas$. We do not show HPMS862 and HPMS3274, their times of closest approach are after J2070.0, and their cutouts can be found through the \href{https://sky.esa.int/}{ESAsky} and \href{http://legacysurvey.org/viewer}{Legacy Surveys}.}
	
	\label{Fig_d+Cutouts}
\end{figure}

\begin{figure}
	\includegraphics[width=\columnwidth]{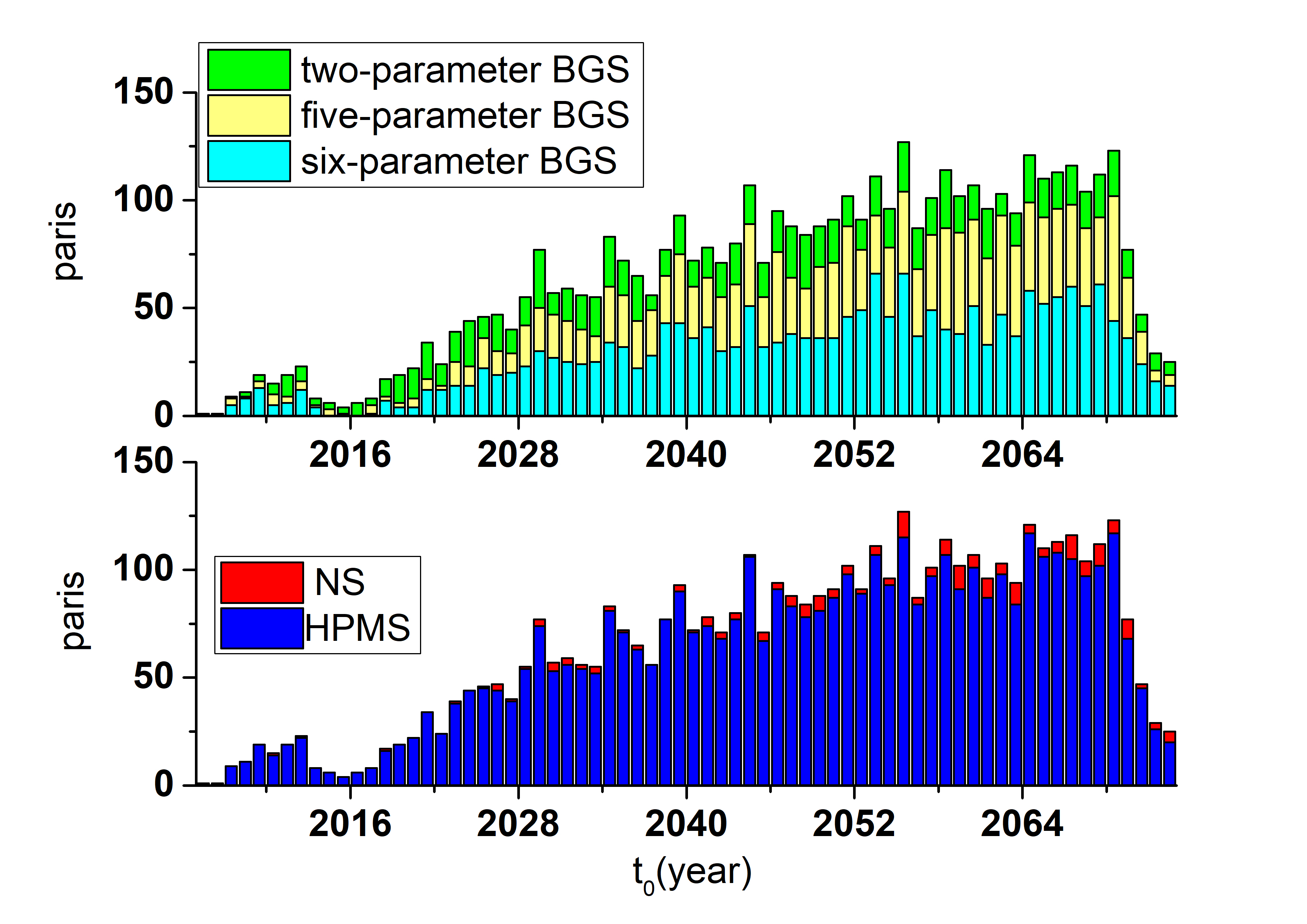}
	\caption{Top panel: the histogram of the closest time for star pairs. The green, yellow and cyan bars show the events with two-, five- and six-parameter solutions BGS, respectively. Bottom panel: the histogram of the closest time for HMPS and NS. The blue and red bars show the events with HMPS and NS lens, respectively. }
	\label{Fig_time_LENS}
\end{figure}

\begin{figure}
	\includegraphics[width=\columnwidth]{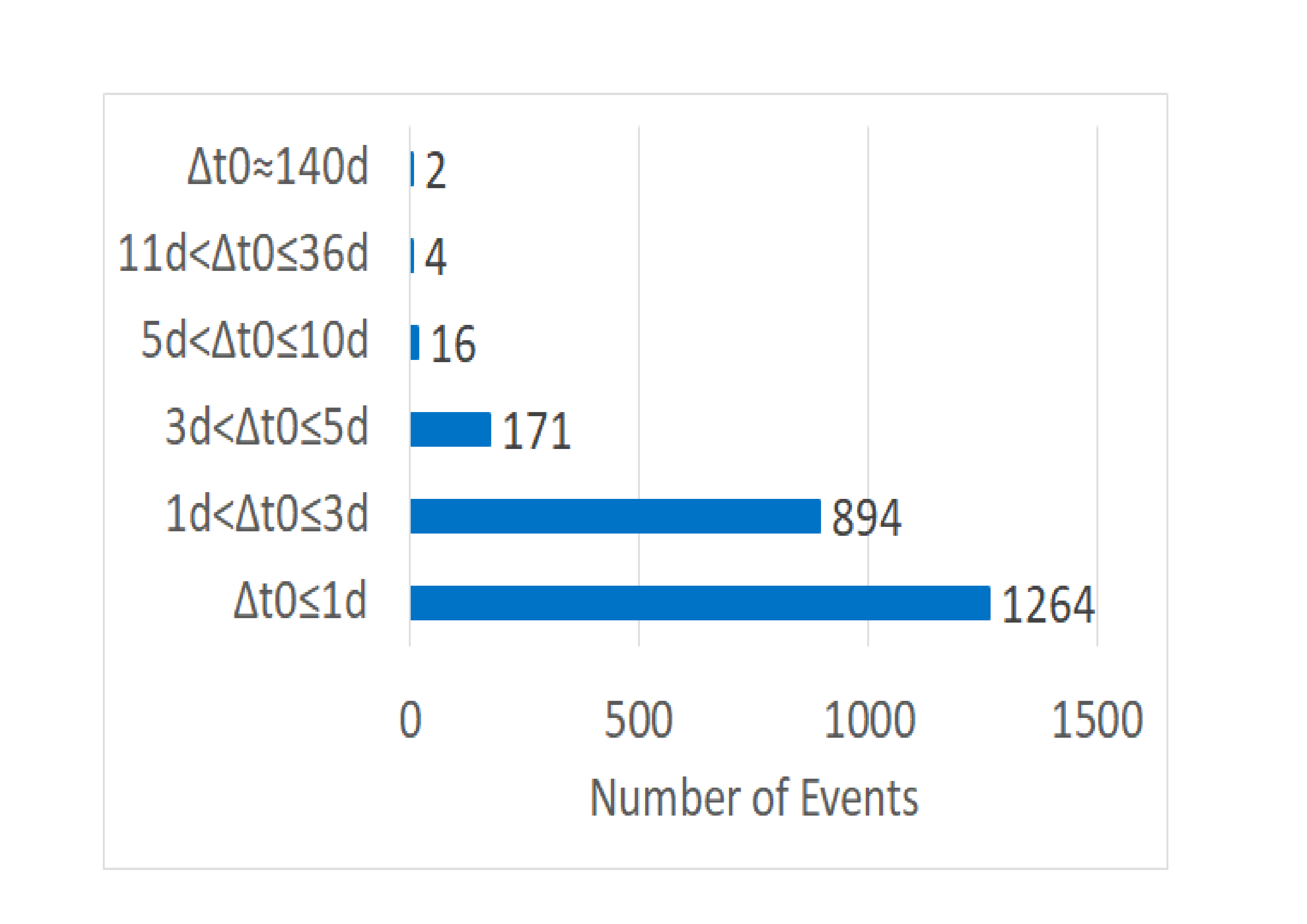}
	\caption{The $\Delta t_0$-statistic diagram of overlap events with the five- or six-parameter background source stars. About 92\% of events have $\Delta t_0 < 3d$.}
	\label{Fig_t0_five}
\end{figure}

\begin{figure*}
	\includegraphics[width=2\columnwidth]{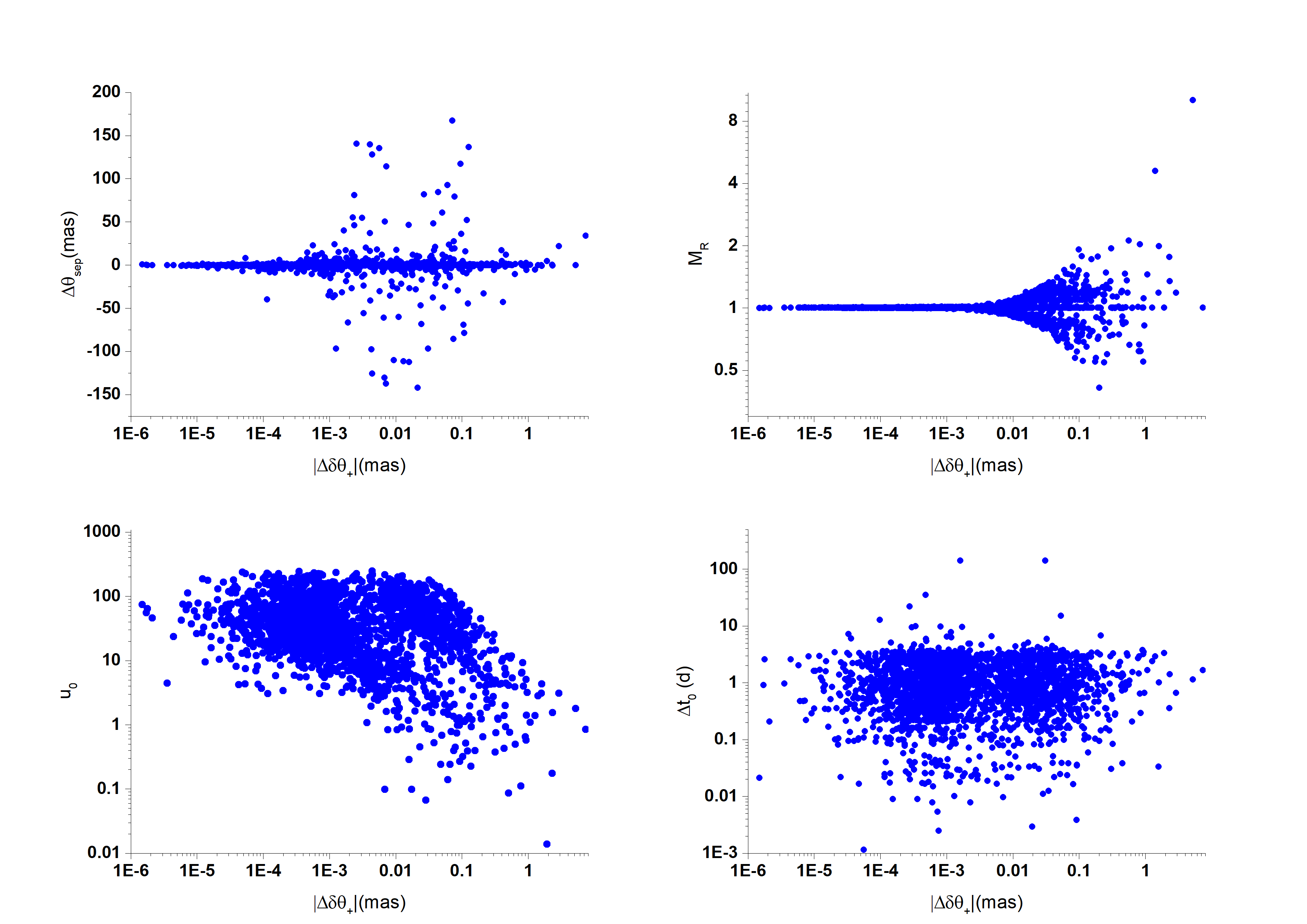}
	\caption{Comparison of $\left |\Delta  \delta\theta_{+}\right |$ for overlap events with the five- or six-parameter background source stars. Upper-left panel: the relationship diagram of  $\left |\Delta  \delta\theta_{+}\right |$ and $\Delta \theta_{sep}$. Upper right panel: the relationship diagram of  $\left |\Delta  \delta\theta_{+}\right |$ and $M_R$. There are four events with $M_R>2$ (HPMS1160,	HPMS3441, HPMS1742 and HPMS1703), and their mass estimates are from the white dwarf catalogue (\citet{Gentile2021}). For the four events, $\Delta \delta\theta_+$ are 5.174 mas, 1.412 mas , 0.827 mas and 0.562 mas, respectively. Lower left panel: the relationship diagram of  $\left |\Delta  \delta\theta_{+}\right |$ and $u_0$. For events with large difference in $\delta\theta_{+}$, $u_0$ is very small. Lower right panel: the relationship diagram of  $\left |\Delta  \delta\theta_{+}\right |$ and $\Delta t_0$. There is not obvious difference in $\delta\theta_+$ for events with large difference of $t_0$. }
	\label{Fig_d+_five}
\end{figure*}

\begin{figure}
	\includegraphics[width=\columnwidth]{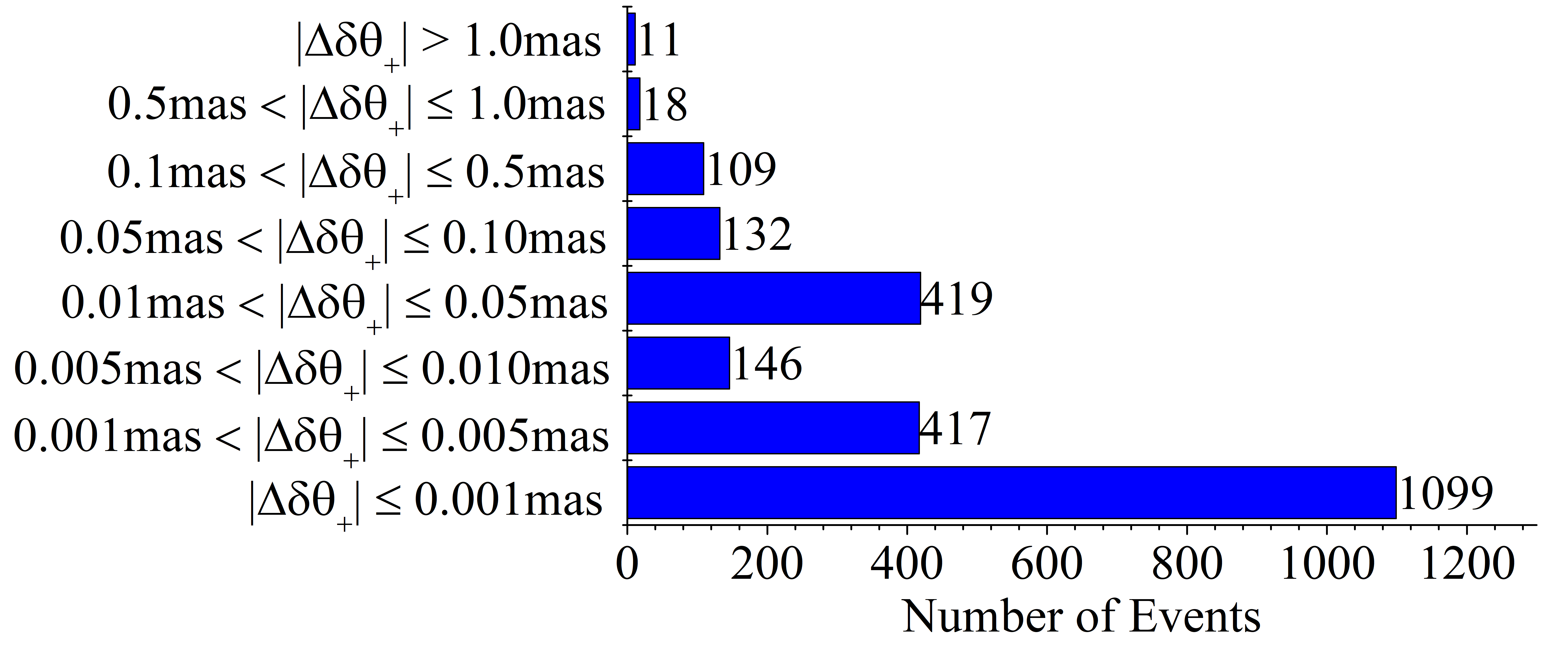}
	\caption{The $\left | \Delta \delta\theta_+ \right |$-statistic diagram of the overlap events, where about 71\% of events have $\left | \Delta \delta\theta_+ \right | \le 0.010 mas$.}
	\label{Fig_five_d+_number}
\end{figure}

\begin{figure*}
	\includegraphics[width=2\columnwidth]{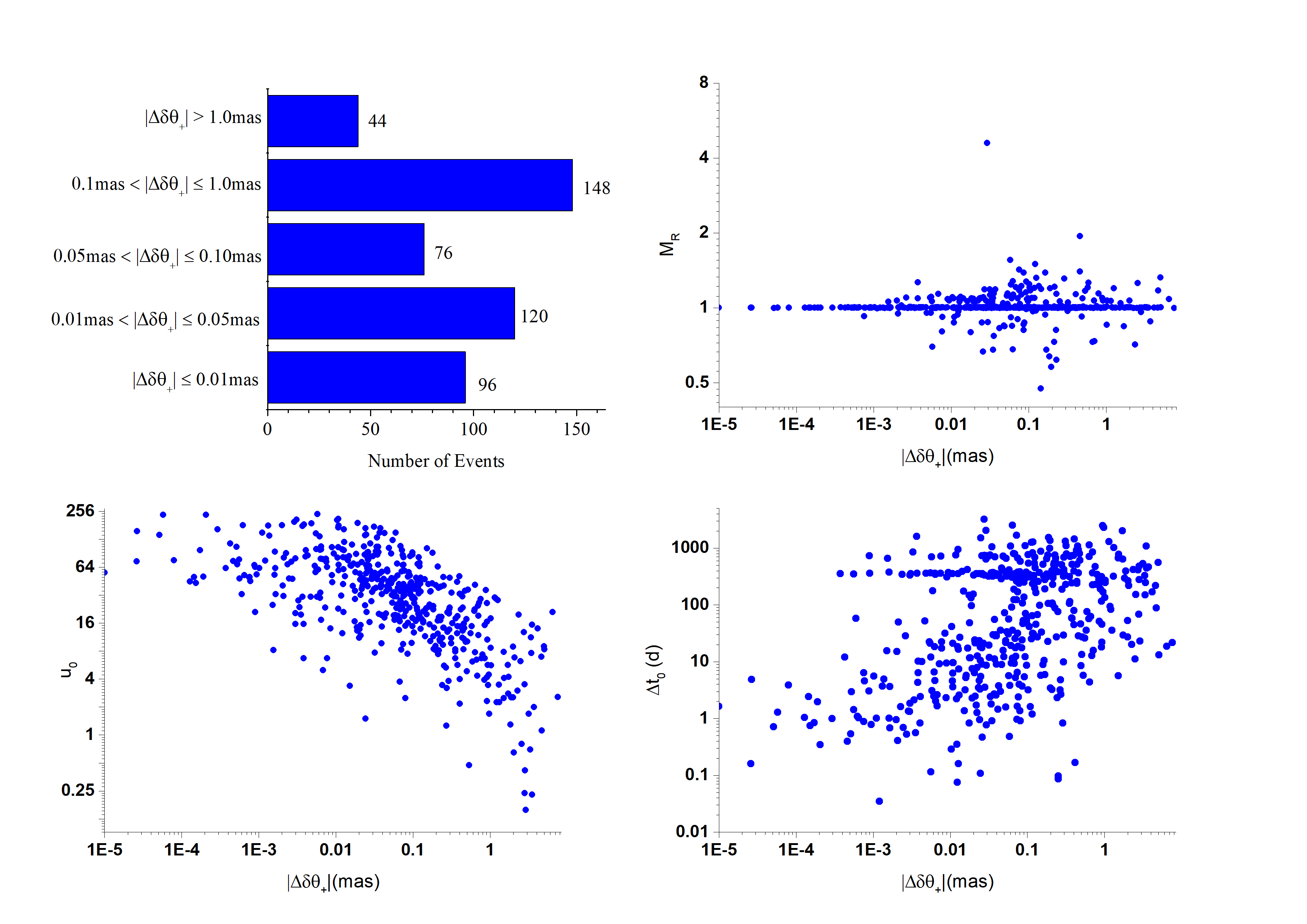}
	\caption{Comparison of $\delta\theta_{+}$ for overlap events with the two-parameter background source stars. Upper-left panel: The $\left | \Delta \delta\theta_+ \right |$-statistic diagram. The contents of other three panels are similar with those of Figure \ref{Fig_d+_five}.}
	\label{Fig_d+_two}
\end{figure*}

\begin{figure}
	\includegraphics[width=\columnwidth]{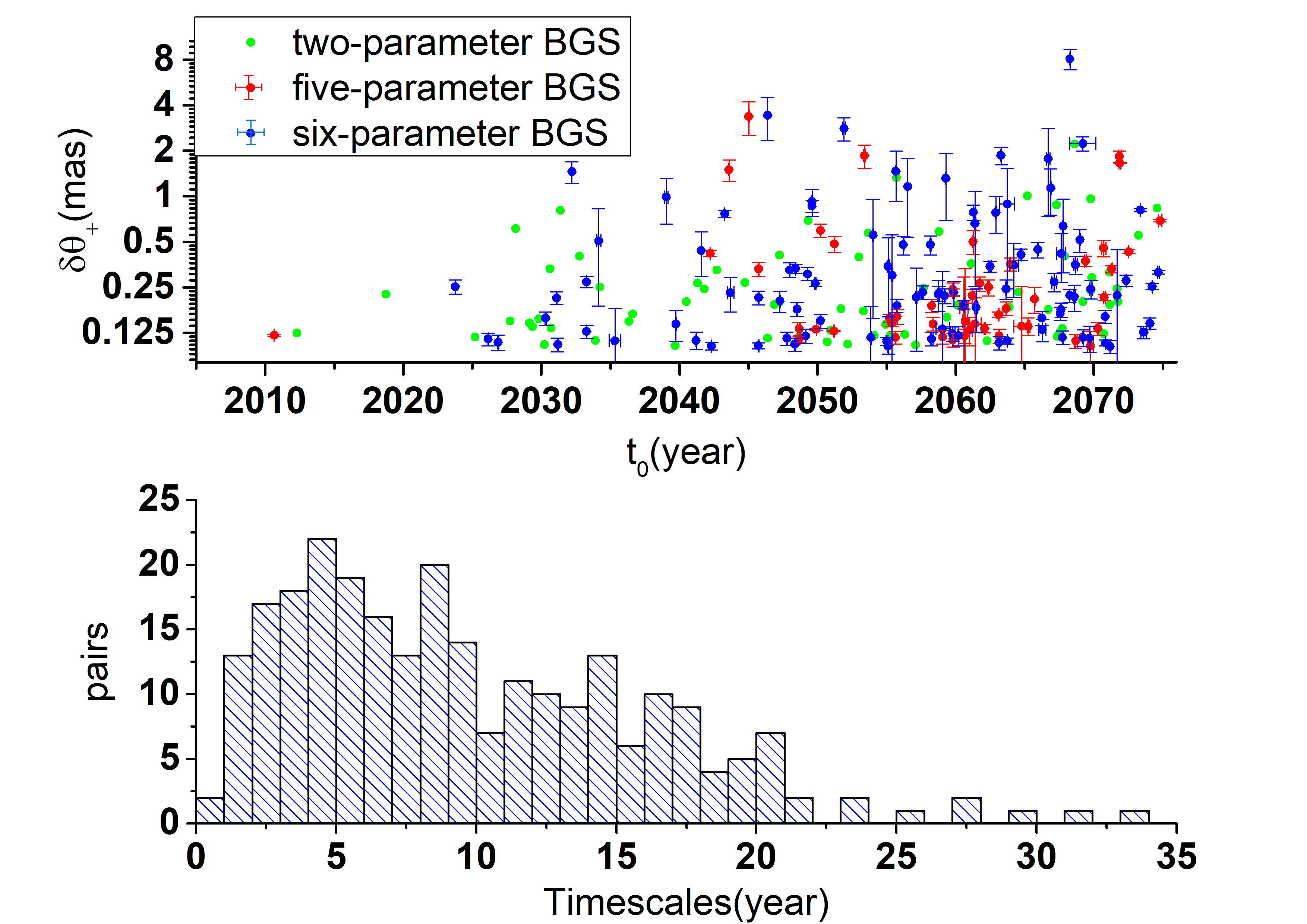}
	\caption{Top panel: the relationship diagram of $t_0$  and $\delta\theta_+$ for microlensing events by NS.The green, red and blue dots show the events with two-, five- and six-parameter solutions BGS, respectively.
		Bottom panel:the histogram of the timescales for star pairs by NS.Four events with a timescale of more than 35 years are not shown in the figure.}
	\label{Fig_big_parallax}
\end{figure}

\begin{figure}
	\includegraphics[width=\columnwidth]{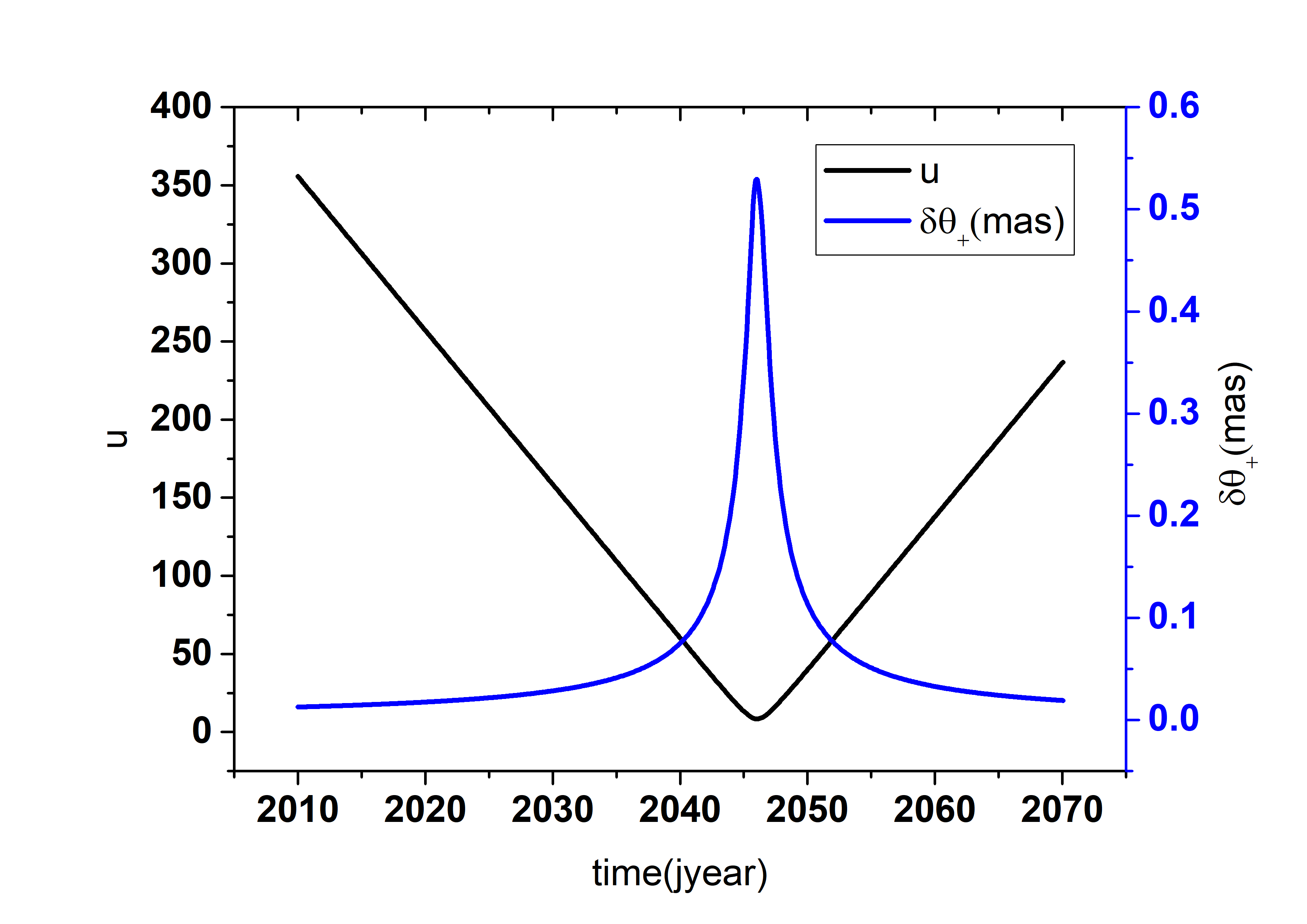}
	\caption{The changes of $u$ (black) and $\delta\theta_+$ (blue) with time for the events HMS1.}
	\label{Fig_bigmass}
\end{figure}

\begin{figure}
	\includegraphics[width=\columnwidth]{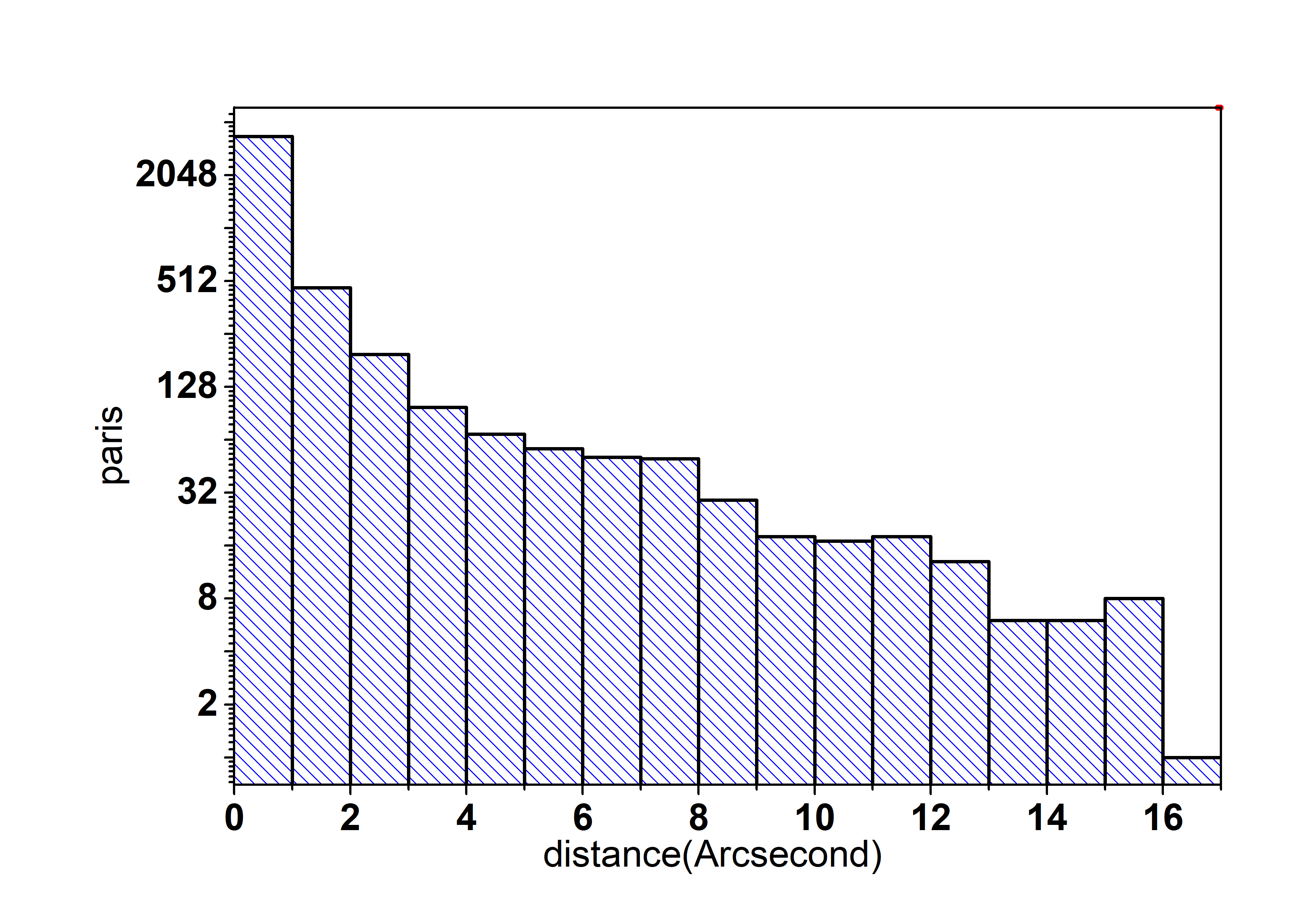}
	\caption{The distances of background stars from the proper motion paths of lens stars for all astrometric microlensing events. There are 116 events with the distances more than 8 arcseconds.}
	\label{Fig_all_distance}
\end{figure}

In this paper, the searching time range is J2010.0-J2070.0, and the result is provided as a spreadsheet available \href{https://osf.io/92p8r/?view\_only=178c791217c644aebc8c125ef0347b92}{online}, where each event has 43 columns and a single row. Table \ref{example} is a sample table consisting of 5 rows of result data. Column meanings in Table \ref{example} are defined in Table \ref{defined}. Finally, 4500 events caused by 3558 lens stars are found, where 4279 events are caused by HPMS (including 907, 1453 and 1919 events with two, five and six astrometric parameter for the background stars, respectively). There are 220 events caused by NS (including 71, 50 and 99 events with two, five and six astrometric parameter for background stars, respectively. Only one event is caused by HMS and has two parameter background star. The details are shown in Table \ref{Statistical results}. In Section \ref{sec:The overlap events}, Section \ref{sec:The events only searched by Klüter} and Section \ref{sec:The events only searched by us}, our results are compared with those of \citet{Kluter22} in detail.

In the paper, the masses of the lens stars are mainly distributed in $M_L<1.2M_{\odot}$ for many events. No event is found for the mass of the lens star distributed in $3.2M_{\odot}<M_L<5M_{\odot}$. For the events found by the authors \citep{Bramich18a,Bramich18b,Kluter18b,Kluter22}, the masses of their lens stars are less than $2.5M_{\odot}$. The mass threshold of HMS is set to be $M_L>5M_{\odot}$, and only one event is finally determined for HMS, more detailed
description is given in Section \ref{sec:The events only searched by us}. Therefore our future work can reduce the mass threshold of lens stars to $3M_{\odot}$ which could help us to find more events.

There are 293 lens stars that can cause two or more events, as shown in Figure \ref{Fig_multiple events}. Surprisingly, there are five lens stars to cause more than 50 events. In section \ref{sec Multiple astrometric microlensing events caused by a single lens star}, we will discuss two of these lens stars and their corresponding events.

About $48\%$ of events have $0.1mas<\delta\theta_{+}<0.2mas$, 457 events have $\delta\theta_{+}>1mas$, and 7 events have $\delta\theta_{+}>10mas$, as shown in Figure \ref{Fig_all_d+} and Figure \ref{Fig_d+Cutouts}. The events around the Gaia DR3 epoch of J2016.0 are the least because of the angular resolution limitation of Gaia DR3. From 2018 to 2035, the events will gradually increase and reach at the rate of about $96 events/yr$, as shown in Figure \ref{Fig_time_LENS}.

\subsubsection{The overlap events}
\label{sec:The overlap events}

There are 2836 events overlapped, we calculate $\Delta t_0=|t_{0,ext}-t_0|$, $\Delta \delta \theta_+=\delta \theta _{+,ext}-\delta \theta _{+}$, $M_R=M_L/M_{L,ext}$ and $\Delta \theta_{sep}=\theta_{sep,ext}-\theta_{sep}$, where the values with subscript 'ext' are the results of \citet{Kluter22} and $\theta_{sep}$ is our estimated distance at closest approach.

There are 2352 events with five or six astrometric
parameter for background stars. For $\Delta t_0$, about 92\% of events have $\Delta t_0 < 3 d$ (as shown in Figure \ref{Fig_t0_five}), but there are two events with $\Delta t_0 \approx 140 d$, where their $\Delta \delta \theta_+$  are very small ($-0.002 mas$ and $- 0.031 mas$ respectively, as shown in lower-right panel of Figure \ref{Fig_d+_five}).  For $\Delta \delta\theta_+$, about 71\% of events have $\left | \Delta \delta\theta_+ \right | \le 0.010 mas$ (as shown in Figure \ref{Fig_five_d+_number}). However, about 5.9 \% of events have $\left | \Delta \delta\theta_+ \right | \ge 0.1 mas$, and we note that $u _ 0$ of these events is relatively small, especially when $\left | \Delta \delta\theta_+ \right | \ge 0.5 mas$, $u _ 0 $ is less than 10 (see lower-left panel of Figure \ref{Fig_d+_five}). It can be seen from  formula (\ref{+}) that $\delta\theta_+$ is related to $u$ and $\theta_E$, and  $\delta\theta_+$ changes rapidly with $u$ when $u$ is small. There are four events with $M_R>2$ (HPMS1160,	HPMS3441, HPMS1742 and HPMS1703), and their mass estimates are from the white dwarf catalogue (\citet{Gentile2021}), $M_L$ are $(0.87\pm 0.09) M_{\odot}$, $(0.44\pm 0.16) M_{\odot}$, $(1.32\pm 0.01) M_{\odot}$ and $(1.38\pm 0.06) M_{\odot}$, respectively. For the four events, $\Delta \delta\theta_+$ are 5.174 mas, 1.412 mas , 0.827 mas and 0.562 mas, respectively (see upper right panel of Figure \ref{Fig_d+_five} ). For these events with $\left | \Delta \delta\theta_+ \right | \ge 0.1 mas$, our estimated lens masses are not the same as that of \citep{Kluter22} (out of which 67\% the mass difference is more than 10\% ) to cause the differences in $\theta _ E$. But, 54 \% of events with $\left | \Delta \delta\theta_+ \right | \ge 0.1 mas$ still have $\delta\theta_+$ differences within $1\sigma$.

For 484 events with two parameter background source stars, the differences of both predicted results are obvious, where the parallax and proper motion of background stars used are different in addition to lens estimated mass.
\citep{Kluter22} used the estimated values of the parallax and proper motion, but we set them to 0. The comparisons of both results are shown in Figure \ref{Fig_d+_two}. It is noted that there are 106 events with $\Delta t_0 < 5 d$ and 215 events with $\left | \Delta \delta\theta_+ \right | \leq 0.05 mas$.

\subsubsection{The events only searched by \citet{Kluter22}}
\label{sec:The events only searched by Klüter}

2006 events in \citet{Kluter22} are not included in our results, where 1578 are excluded because they do not meet the selecting conditions in Table \ref{conditions_lens}, Table \ref{conditions_background}, or the equation (\ref{d_parallax}). There are 166 events that do not satisfy the selection of two-parameter background stars in Section \ref{sec:Checking for microlensing events}. In addition, we find that $\delta\theta_{+}$ of 262 events do not reach the threshold of 0.1 mas. These differences could be caused by lens mass estimation, the parallax and proper motion of background stars. See Table \ref{Exclusion reasons1} for details. However, $\sim 80\%$ of these 262 events have a difference of $\delta\theta_{+}$ within $1\sigma$.

\subsubsection{The events only searched by us (the new predicted event)}
\label{sec:The events only searched by us}

We have searched 1664 new predicted event, where about 86 \% of the events are not included in the search range of \citet{Kluter22}, including the lens, background star and $t_0$ of event. It is noted that the sources of some events are not within the rectangular box set by \citet{Kluter22} during the initial search. Other events could be related to the differences of selecting parameters, such as lens mass estimation, the parallax and proper motion of background stars. See Table \ref{Exclusion reasons2} for details.

We will discuss these new events in detail based on our classification of lens.

(1) Microlensing events by NS

There are 220 events caused by NS (see Table \ref{Statistical results}), where 45 events with $ \delta\theta_+ >0.5 mas$, out of which 20 events  with $ \delta\theta_+ >1  mas$ (including 4, 5 and 11 events with two, five and six parameter BGS, respectively). 27 events will occur between J2010.0 and J2035.0, of which 2 events with $ \delta\theta_+ >1  mas$. After J2035.0, the number of events will increase, about 5 events per year (see Figure \ref{Fig_time_LENS} and Figure \ref{Fig_big_parallax}). For timescales of the events with $ \delta\theta_+ >0.1  mas$, 62 events are within 5 years, 67 events are between 5 years and 10 years, 91 events are longer than 10 yr (see Figure \ref{Fig_big_parallax}).

NS180. This event pick will occur in $2032.21 \pm 0.03$, the timescale is $\sim 6.2 yr$ and the maximum shift of the major image is $(1.45\pm 0.24) mas$. The lens mass is estimated to be $0.32 M_{\odot}$ by the mass-luminosity relations in the third case of Section \ref{sec:Mass estimation of lens stars}. Its total proper motion and parallax are $95.777 mas/yr$ and $11.334 mas$, respectively. The lens has $ G\approx 14.77 mag$ which is $\sim 2.78 mag$ brighter than the source star with six parameter solutions. This event is listed in the first row of Table \ref{example}.

(2) Microlensing events by HMS

Initially, for HMS we obtained 558 pairs with $ \delta\theta_+ >0.1  mas$ (including 4 and 554 events with two, five or six parameter BGS, respectively). For 554 events with five or six parameter BGS, 487 pairs can not meet the conditions of table \ref{conditions_background} (fidelity\_v2>0.7), and the remaining 67 pairs did not pass the criteria in equation \ref{d_parallax}. For 4 events with two parameter BGS, 3 events have bad match with Gaia DR2.In the end, we only get an event (HMS1) that satisfies all the conditions.

HMS1. The lens star (source\_id=4090028395985895552) is a single star, its spectral type is B and its mass is estimated as $5.160^{+0.098}_{-0.089} M_{\odot}$,taken from  Gaia DR3 astrophysical parameters. Its parallax, proper motion in right ascension direction and declination direction are $(0.489 \pm 0.017) mas$, $(-22.466 \pm 0.018) mas/yr$ and $(-38.676 \pm 0.015) mas/yr$, respectively.  The lens has $ G\approx 12.77 mag$ which is $\sim 3.1 mag$  brighter than the source star. The event has a time of closest lens-source approach of $ 2046.0 \pm 3.3 yr$ and the Einstein radius of $ (4.54 \pm 3.19) mas$. 
The maximum shift of the major image is $0.53 mas$ and its lower confidence level (16\%) is $0.05 mas$. The changes of $u$ and $\delta\theta_+$ with time is shown in Figure \ref{Fig_bigmass}, the timescale is $\sim 9 yr$. More details are shown in the second row of Table \ref{example}.

(3) Microlensing events by HPMS

We find 1443 new events by HPMS,where 240 events with $ \delta\theta_+ >0.5 mas$, out of which 123 events  with $ \delta\theta_+ >1  mas$ (including 83 events with five or six parameter BGS). There are 845 events occurred between J2010.0 and J2067.0, of which 100 will occur within the next 10 years. In 116 events, the distance of background stars from the proper motion track of the lens star exceeds $8^{\prime\prime}$, as shown in Figure \ref{Fig_all_distance}, where the maximum is $16^{\prime\prime}$, and $\sim 97\%$ of these events have a Einstein radius of more than 30 mas, with a maximum of $\sim 40.66 mas$, so these lens can produce microgravitational effects with $\delta\theta_+>0.1 mas$ on the background stars far away, and $\sim 83\%$ of these lens can cause ten or more events because of high proper motion. However, since $\delta\theta_+$ decreases with the increase of $\theta_{sep}$,  the expected shift is small ($\delta\theta_+ < 0.2 mas$).

There are three examples listed in Table \ref{example}:

HPMS1816. The maximum shift of the major image is more than 10 mas ($\delta\theta_+= 10.94 mas$) and its lower confidence level (16\%) is $0.79 mas$. The event has a time of closest lens-source approach of $2068.48 \pm 1.62 yr$. The lens (CD-29 5220) has $ G\approx 9.95 mag$ which is $\sim 11 mag$  brighter than the source star and the source star only have two parameter solutions. The lens mass is estimated to be $0.65 M_{\odot}$ by the mass-luminosity relations in the third case of Section \ref{sec:Mass estimation of lens stars}. Its total proper motion and parallax are $162.919 mas/yr$ and $23.662 mas$, respectively. 

HPMS3274. The maximum shift of the major image is the largest of all events ($\delta\theta_+= 21.38 mas$) and its lower confidence level (16\%) is $4.82 mas$. The Einstein radius is $(30.54 \pm 0.89) mas$. However, the event peak will occur in $2073.23 \pm 0.23 $ and the lens (V$^*$ V2215 Oph) is very bright ($ G\approx 5.89 mag$) which is $\sim 14.87 mag$ brighter than the source star with two parameter solutions. The lens star is a single star, its spectral type is K and its mass is estimated as $0.682^{+0.040}_{-0.040} M_{\odot}$, taken from  Gaia DR3 astrophysical parameters.

HPMS4600. The maximum shift of the major image is $(6.72\pm 2.17) mas$, the event peak will occur in $2069.47 \pm 0.12$ and the timescale is $\sim 6.4 yr$. The lens mass is estimated as $(0.42\pm 0.02) M_{\odot}$ from the white dwarf catalogue (\citet{Gentile2021}) labeled as 'WDJ221800.59+560214.92'. The lens with $ G\approx 18.03 mag$ is $\sim 2.44 mag$ brighter than the source star with six parameter solutions.

\begin{table*}
	\caption{ 5 rows of the predicted astrometric-microlensing events in this paper, and the rest are available \href{https://osf.io/92p8r/?view\_only=178c791217c644aebc8c125ef0347b92}{online}.}
	\label{example}
	\begin{tabular}{ccccccccc}
		\hline
		\hline
		Event\_name & ref\_epoch& lens\_source\_id & lens\_ra & lens\_dec & lens\_parallax & lens\_pm& lens\_pmra&	lens\_pmdec  \\
		-&yr&-&deg&deg&mas&mas/yr&mas/yr&mas/yr\\
		\hline
		NS180    & 2016 & 5255174721885285248 & 157.677070264 & -59.942343487 & 11.334  & 95.777   & -80.225  & -52.318   \\
		HMS1     & 2016 & 4090028395985895552 & 277.984338089 & -22.029599100 & 0.489   & 44.727   & -22.466  & -38.676   \\
		HPMS1816 & 2016 & 5597168291724032512 & 119.328289919 & -30.147903097 & 23.662  & 162.919  & -15.912  & -162.140  \\
		HPMS3274 & 2016 & 4109034455276324608 & 259.053294100 & -26.551146095 & 167.962 & 1222.339 & -479.573 & -1124.332 \\
		HPMS4600 & 2016 & 2006217676803960960 & 334.503341666 & 56.038507792  & 25.406  & 256.966  & 111.256  & 231.632 \\				  
		\hline	
		\multicolumn{9}{l}{Notes. NS180 pick will occur in 2032, the timescale is $\sim 6.2 yr$, the maximum shift of the major image is $(1.45\pm 0.24) mas$,}\\
		\multicolumn{9}{l}{For HMS1, the lens estimation mass  is more than $5 M_{\odot}$.}\\
		\multicolumn{9}{l}{For HPMS1816 and HPMS3274, the maximum shift of the major image are more than 10 mas.}\\
		\multicolumn{9}{l}{For HPMS4600, the lens is White Dwarf, the maximum shift of the major image is $(6.72\pm 2.17) mas$.}\\
	\end{tabular}
\end{table*}

\begin{table*}
	\contcaption{5 rows of the predicted astrometric-microlensing events in this paper, and the rest are available \href{https://osf.io/92p8r/?view\_only=178c791217c644aebc8c125ef0347b92}{online}.}
	\label{tab:continued}
	\begin{tabular}{cccccccc}
		\hline
		\hline
		lens\_g\_mean &lens\_M&	WDJ\_name&	Pwd	&type&	lens\_fidelity\_v2&	source\_source\_id&	source\_ra\\
		mag & solMass&-&-&-&-&-&deg \\
		\hline
		14.774 & 0.32 & -                      & -     & MS    & 1.000 & 5255174721854095872 & 157.676416918 \\
		12.775 & 5.16 & -                      & -     & -     & 1.000 & 4090028400306349568 & 277.984125875 \\
		9.952  & 0.65 & -                      & -     & MS    & 1.000 & 5597168296018750080 & 119.328018818 \\
		5.890  & 0.68 & -                      & -     & -     & 1.000 & 4109028510993111040 & 259.044797669 \\
		18.034 & 0.42 & WDJ221800.59+560214.92 & 0.985 & WD & 1.000 & 2006218535799265024 & 334.506387815  \\
		\hline		
	\end{tabular}
\end{table*}

\begin{table*}
	\contcaption{5 rows of the predicted astrometric-microlensing events in this paper, and the rest are available \href{https://osf.io/92p8r/?view\_only=178c791217c644aebc8c125ef0347b92}{online}.}
	\label{tab:continued}
	\begin{tabular}{cccccccc}
		\hline
		\hline
		source\_dec&	source\_parallax& source\_pm&	source\_pmra&	source\_pmdec&	source\_g\_mean&	source\_ruwe(new)&	source\_fidelity\_v2\\
		deg&mas&mas/yr&	mas/yr&	mas/yr&	mag&-&- \\
		\hline
		-59.942597575 & 0.274 & 9.248 & -7.977 & 4.680  & 17.554 & -   & 0.832 \\
		-22.029916621 & 0.000 & 0.000 & 0.000  & 0.000  & 15.878 & 1.4 & -     \\
		-30.150261708 & 0.000 & 0.000 & 0.000  & 0.000  & 20.965 & 0.8 & -     \\
		-26.569040325 & 0.000 & 0.000 & 0.000  & 0.000  & 20.761 & 1.1 & -     \\
		56.041993407  & 1.078 & 4.041 & -3.024 & -2.681 & 20.476 & -   & 0.987 \\
		\hline		
	\end{tabular}
\end{table*}

\begin{table*}
	\contcaption{5 rows of the predicted astrometric-microlensing events in this paper, and the rest are available \href{https://osf.io/92p8r/?view\_only=178c791217c644aebc8c125ef0347b92}{online}.}
	\label{tab:continued}
	\begin{tabular}{llllllllllllll}
		\hline
		\hline
		sep   & e\_sep & E     & e\_E & u0   & e\_u0 & d+    & e\_d+ & d+\_lower & dC    & e\_dC & dC\_lower & dC\_lum & e\_dC\_lum \\
		mas   & mas    & mas   & mas  & -    & -     & mas   & mas   & mas       & mas   & mas   & mas       & mas     & mas        \\ \hline
		18.53 & 2.78   & 5.38  & 0.27 & 3.44 & 0.55  & 1.45  & 0.24  & -         & 1.34  & 0.18  & -         & 0.11    & 0.02       \\
		38.32 & 91.18  & 4.54  & 3.19 & 8.45 & 52.11 & 0.53  & -     & 0.05      & 0.52  & -     & 0.04      & 0.03    & -          \\
		0.50  & 153.89 & 11.18 & 0.60 & 0.04 & 14.36 & 10.94 & -     & 0.79      & 3.95  & -     & 0.78      & 0.00    & -          \\
		22.25 & 166.31 & 30.54 & 0.89 & 0.73 & 5.52  & 21.38 & -     & 4.82      & 10.80 & -     & 4.70      & 0.00    & -          \\
		5.73  & 19.37  & 9.14  & 0.26 & 0.63 & 2.13  & 6.72  & 2.17  & -         & 3.23  & 0.78  & -         & 0.77    & 0.22  \\ \hline    
	\end{tabular}
\end{table*}

\begin{table*}
	\contcaption{5 rows of the predicted astrometric-microlensing events in this paper, and the rest are available \href{https://osf.io/92p8r/?view\_only=178c791217c644aebc8c125ef0347b92}{online}.}
	\label{tab:continued}
\begin{tabular}{llll}
	\hline
	\hline
	dC\_lum\_lower & t0        & e\_t0  & new events \\
	mas            & yr        & yr     &            \\ \hline
	-              & 2032.2067 & 0.0277 & *          \\
	0.00          & 2046.0000 & 3.3156 & *          \\
	0.00           & 2068.4830 & 1.6180 & *          \\
	0.00           & 2073.2950 & 0.2298 & *          \\
	-              & 2069.4857 & 0.1241 & *   \\ \hline      
\end{tabular}
\end{table*}

\begin{table*}
	\caption{Column meanings in Table \ref{example} .}
	\label{defined}
	\begin{tabular}{cll}
		\hline
		\hline
		Column Number  & Column Name  & Description      \\
		\hline
		1   & Event\_name  &The name of the predicted astrometric microlensing event.   \\
		2   & ref\_epoch  & Reference epoch  \\
		3   & lens\_source\_id & Gaia DR3 source\_id of the lens   \\
		4   & lens\_ra  & Right ascension of the lens   \\
		5   & lens\_dec  & Declination of the lens  \\
		6   & lens\_parallax  & Parallax of the lens\\
		7   & lens\_pm & Total proper motion  of the lens \\
		8   & lens\_pmra  & Proper motion in right ascension direction of the lens   \\
		9   & lens\_pmdec  & Proper motion in declination direction of the lens \\
		10   & lens\_g\_mean  & G-band mean magnitude of the lens  \\
		11   & lens\_M  & The estimated mass of the lens  \\
		12   &WDJ\_name  & The name of a white dwarf from 'the white dwarf catalog' \citep{Gentile2021}  \\
		13   & Pwd  &the probability of the source becoming a white dwarf from \citet{Gentile2021} \\
		14   & type  & \begin{tabular}[l]{@{}l@{}} Type of the lensing star: WD = White Dwarf,  MS = Main Sequence, RG = Red Giant,\\ BD = Brown Dwarf\end{tabular}\\
		15   & lens\_fidelity\_v2& Astrometric fidelity of the lens from \citet{Rybizki22}  \\
		\hline
		16   & source\_source\_id &  Gaia DR3 source\_id of the background source stars \\
		17   & source\_ra &  Right ascension of the background source stars \\
		18   &source\_dec  &  Declination of the background source stars \\
		19  &source\_parallax& Parallax of the background source stars\\
		20 &source\_pm& Total proper motion  of the background source stars\\
		21  &source\_pmra&  Proper motion in right ascension direction of the background source stars\\
		22  & source\_pmdec&  Proper motion in declination direction of the background source stars \\
		23  &source\_g\_mean&   G-band mean magnitude of the background source stars  \\
		24  &source\_ruwe(new)& \makecell[l]{ Renormalised unit weight error for the background source star with two-parameter solution,\\and can be solved by formula(\ref{ruwe})}\\
		
		25  &source\_fidelity\_v2&  Astrometric fidelity of the background source stars from \citet{Rybizki22} \\
		\hline
		26  &sep& Estimated distance at closest approach\\
		27 &e\_sep&  Error in sep\\
		28  &E&  Einstein radius of the event\\
		29  &e\_E& Error in Einstein radius of the event\\
		30  &u0&  Estimated distance at closest approach in Einstein radii	\\
		31  &e\_u0&  Error in u0\\
		32	&d+&   Maximal astrometric shift of brighter image	\\
		33  &e\_d+&	 Error in d+\\
		34  & d+\_lower&  Lower confidence level (16\%) of d+       \\
		35   &dC&    Maximal astrometric shift of center of light\\
		36 	&e\_dC&  Error in dC\\
		37  & dC\_lower & Lower confidence level (16\%) of dC \\
		38	&dC\_lum&   Maximal astrometric shift including lens-luminosity effects\\
		39	&e\_dC\_lum&	Error in dC\_lum\\
		40  & dC\_lum\_lower & Lower confidence level (16\%) of dC\_lum \\
		41  &t0&   Estimated time of the closest approach\\
		42	&e\_t0&  Error in t0\\
		43 & new events & Marked as * represents a new event different from \citet{Kluter22}\\		
		\hline	
	\end{tabular}
\end{table*}

\begin{table}
	\centering
	\caption{Statistical results of events caused by three types of lens stars and their corresponding background stars. }
	\label{Statistical results}
	\begin{tabular}{|c|c|c|c|c|}
		\hline
		\hline
		& solutions                & HPMS & NS  & HMS \\ \hline
		\multirow{3}{*}{BGS} & two-parameter solutions  & 907 & 71 & 1   \\ \cline{2-5}
		& five-parameter solutions & 1453 & 50  & -   \\ \cline{2-5}
		& six-parameter solutions  & 1919 & 99 & -   \\ \hline
	\end{tabular}
\end{table}

\begin{table}
	\centering
	\caption{Reasons why the results in \citet{Kluter22} are not included in this paper. }
	\label{Exclusion reasons1}
	\begin{tabular}{ccc}
		\hline
		\hline
 & \multicolumn{2}{c}{BGS}                                                                                                                         \\ \hline
Exclusion reasons                                                                           & \begin{tabular}[c]{@{}c@{}}two-parameter \\ solutions\end{tabular} & \begin{tabular}[c]{@{}c@{}}five or\\ six-parameter \\ solutions\end{tabular} \\   \hline
Lens fidelity \_ v2 \textless 0.8                                                           & -                                                                  & 2                                                                          \\
BGS fidelity \_ v2 \textless 0.7                                                            & -                                                                  & 1560                                                                       \\
No satisfying equations (\ref{d_parallax})                                                                        & -                                                                    & 16                                                                         \\
\begin{tabular}[c]{@{}c@{}}No matching with DR2 \\ or external catalogues.\end{tabular} & 166                                                                & -                                                                          \\
$\delta_{+}<0.1mas$  &173                                                                & 89                                                                   	    \\ \hline
				
	\end{tabular}
\end{table}

\begin{table}
	\centering
	\caption{The reason why the results of this paper are not included in \citet{Kluter22}. }
	\label{Exclusion reasons2}
	\begin{tabular}{cccc}
		\hline
		\hline
		&                 & \multicolumn{2}{c}{BGS}                                                                                                                       \\ \hline
		Reasons                                                                             &                 & \begin{tabular}[c]{@{}c@{}}two-parameter\\  solutions\end{tabular} & \begin{tabular}[c]{@{}c@{}}five or\\ six-parameter\\ solutions\end{tabular} \\ \hline
		\multirow{4}{*}{\begin{tabular}[c]{@{}c@{}}Not in the \\ search range\end{tabular}} & lens            & 159                                                               & 473                                                                       \\
		& BGS            & 0                                                               & 12                                                                       \\
		& time            & 109                                                               & 484                                                                       \\
		& rectangular box & 41                                                                & 119                                                                       \\ \hline
		Data Differences                                                                    &                 & 186                                                               & 81   \\ \hline
	\end{tabular}
\end{table}

\subsection{Multiple astrometric microlensing events caused by a single lens star}
\label{sec Multiple astrometric microlensing events caused by a single lens star}

\begin{figure*}
	\includegraphics[width=2\columnwidth]{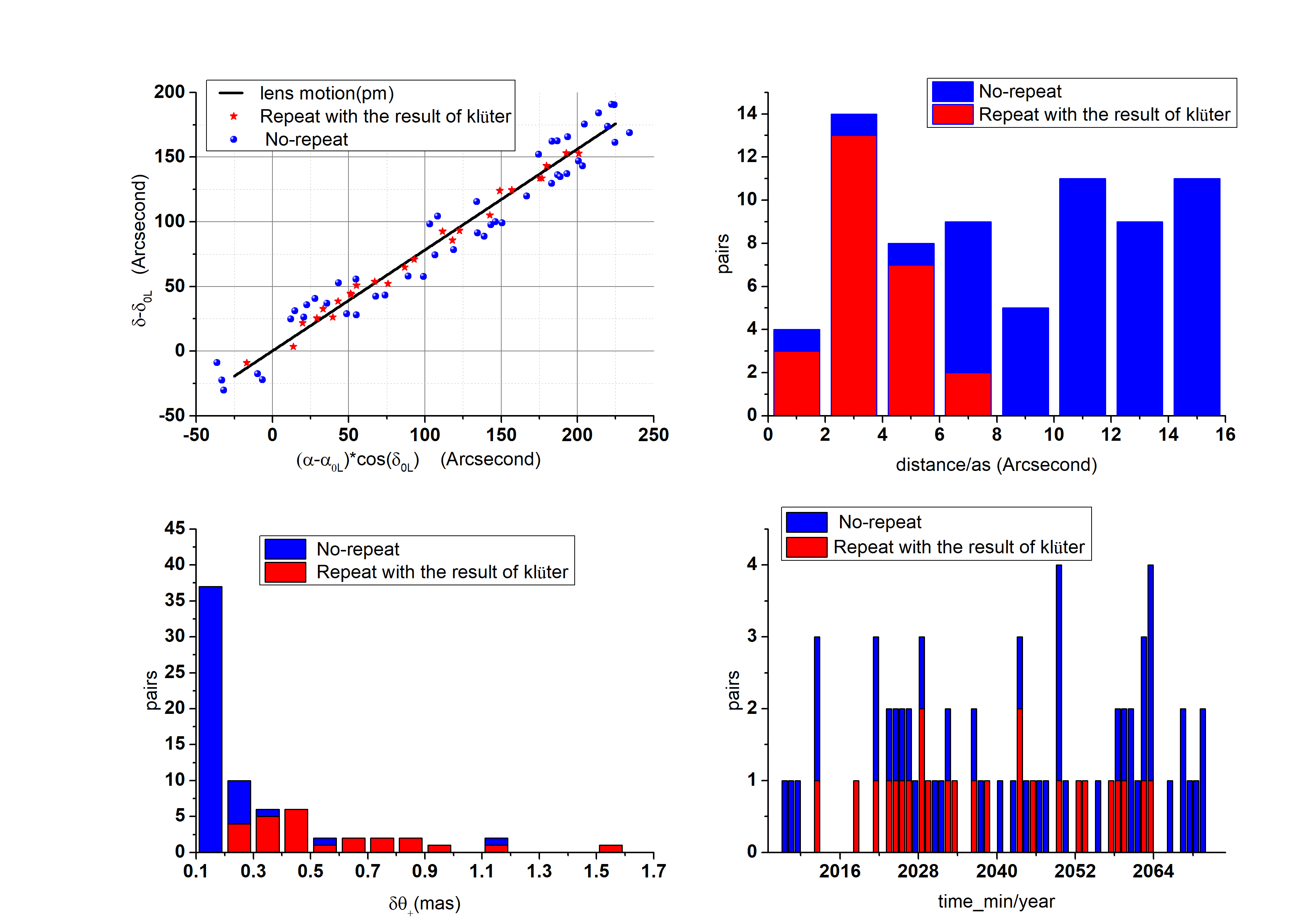}
	\caption{The information of 71 astrometric microlensing events caused by 61 CYG A. Upper-left panel: the distance of background stars in the reference epoch from the proper motion path of lens stars. The black line is the proper motions of the lens star in J2010.0-2070.0, the red asterisks are the background stars (reference epoch time position) in the result of \citet{Kluter22}, and the blue dots are the background star (reference epoch time position) that are not present in the result of \citet{Kluter22}. Upper right panel: the histogram of the distance between the background star and the proper motion track of the lens star. Lower left panel: the histogram of $\delta\theta_{+}$ for 71 events, with 3 events $\delta\theta_{+}>1mas$, and the maximum value of $\delta\theta_{+}$ is 1.51mas. Lower right panel: the histogram of $t_0$.}
	\label{Fig_background78}
\end{figure*}

\begin{figure}
	\includegraphics[width=\columnwidth]{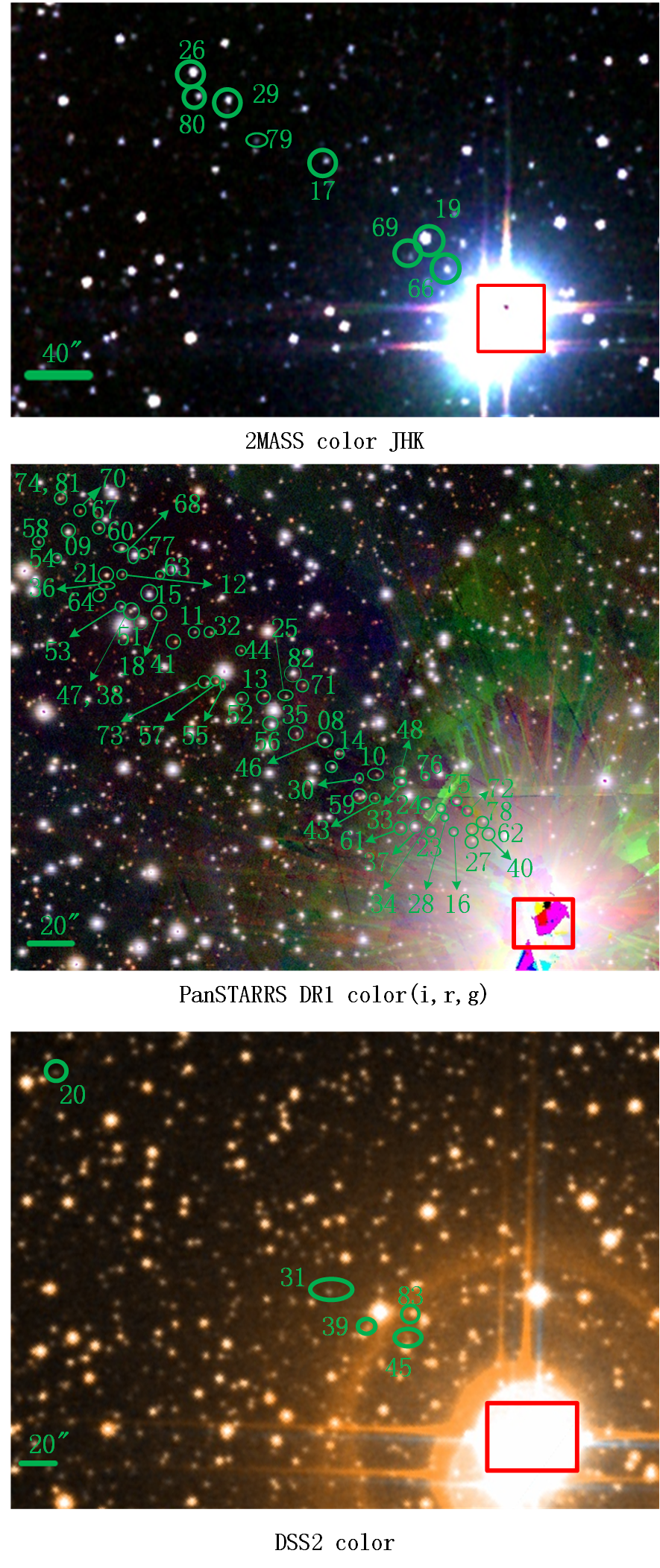}
	\caption{Cutouts for 61 CYG A and its source stars (70) are obtained using the  \href{https://sky.esa.int/}{ESAsky} online observation tool.The source of the HPMS4722 is not marked in these images because it is too close to HPMS4717 and dim. The lens are indicated as red rectangular, and the sources are indicated as green circles."HPMS47" + number (next to the source star) is the event name. Top panel: 2MASS color JHK image (epoch $\sim$ J2000) with 8 events.Middle panel:PanSTARRS DR1 color(i,r,g) image (epoch $\sim$ J2012) with 57 events.Bottom panel:DSS2 color image (epoch $\sim$ J1990) with 5 events.}
	\label{Fig_lens2}
\end{figure}

\begin{figure*}
	\includegraphics[width=2\columnwidth]{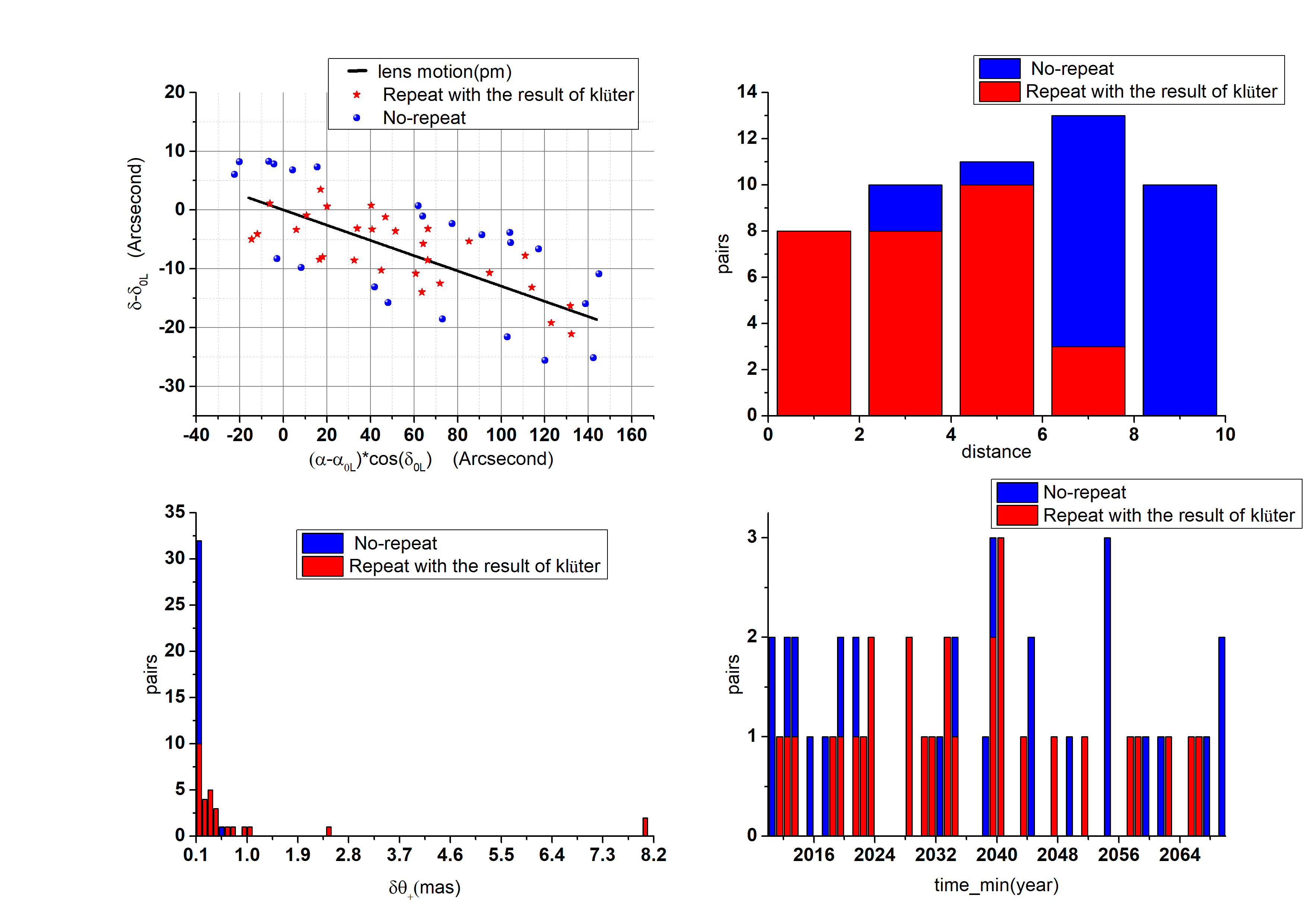}
	\caption{The information of 52 astrometric microlensing events caused by LAWD 37. The format of the figure is the same as in Fig. \ref{Fig_background78}}
	\label{Fig_background77}
\end{figure*}

\begin{figure}
	\includegraphics[width=\columnwidth]{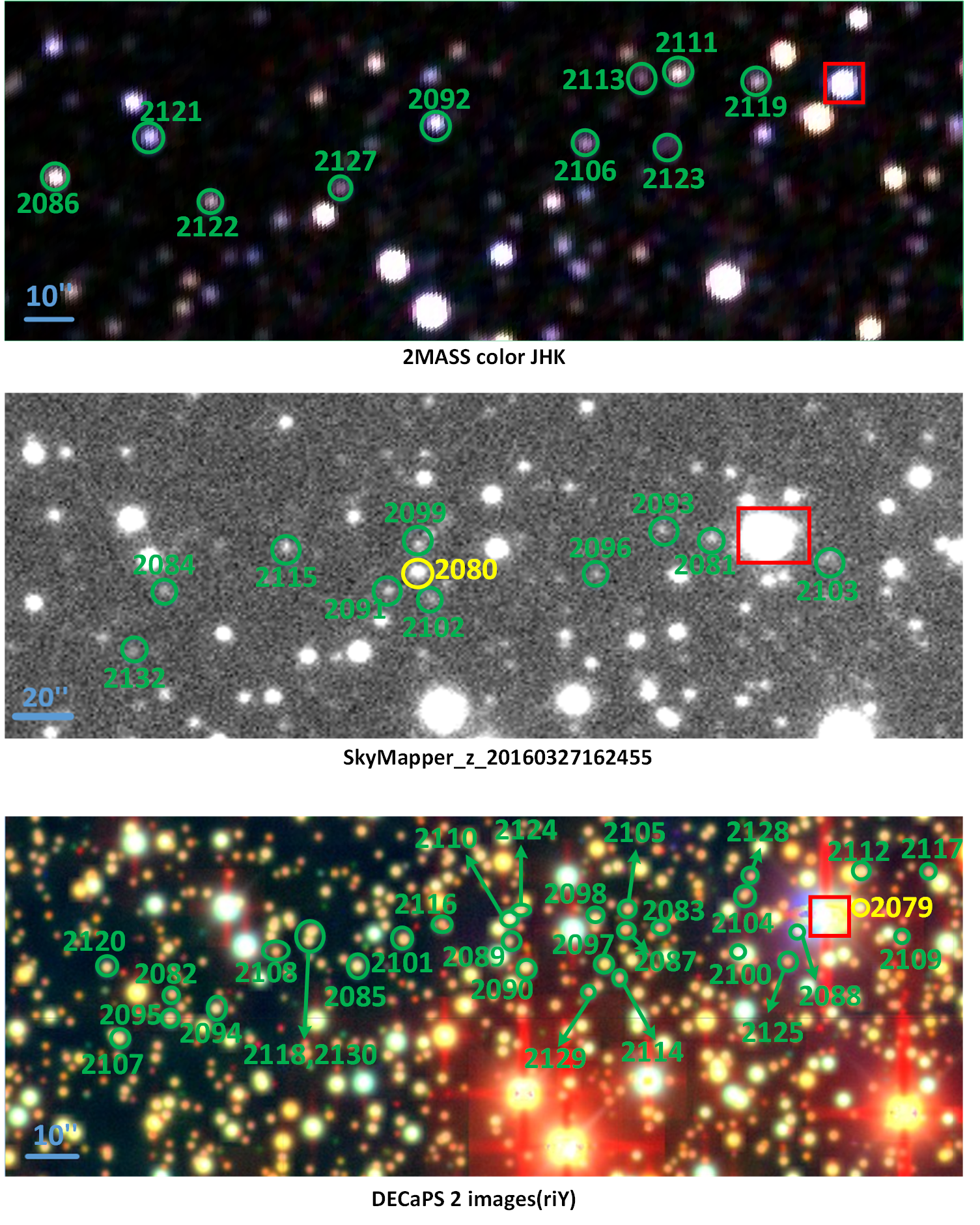}
	\caption{Cutouts for LAWD 37  and its source stars (52).The lens are indicated as red rectangular, the sources are indicated as green circles, HPMS2079 and HPMS2080 ($\delta\theta_{+}>8 mas$) is indicated as yellow circles. "HPMS" + number (next to the source star) is the event name. Top panel: 2MASS color JHK image (epoch $\sim$ J2000) with 10 events are obtained using the \href{https://sky.esa.int/}{ESAsky} online observation tool.Middle panel:SkyMapper image (epoch $\sim$ J2016) with 11 events are obtained through the website  \href{https://skymapper.anu.edu.au/image-cutout/}{SkyMapper}\citep{SkyMapper2016,SkyMapper2018,SkyMapper2019}.Bottom panel:DECaPS 2 images(riY)(epoch $\sim$ J2017) with 31 events are obtained using the \href{http://legacysurvey.org/viewer}{Legacy Surveys} online observation tool. }
	\label{Fig_lens1}
\end{figure}

\begin{figure}
	\includegraphics[width=\columnwidth]{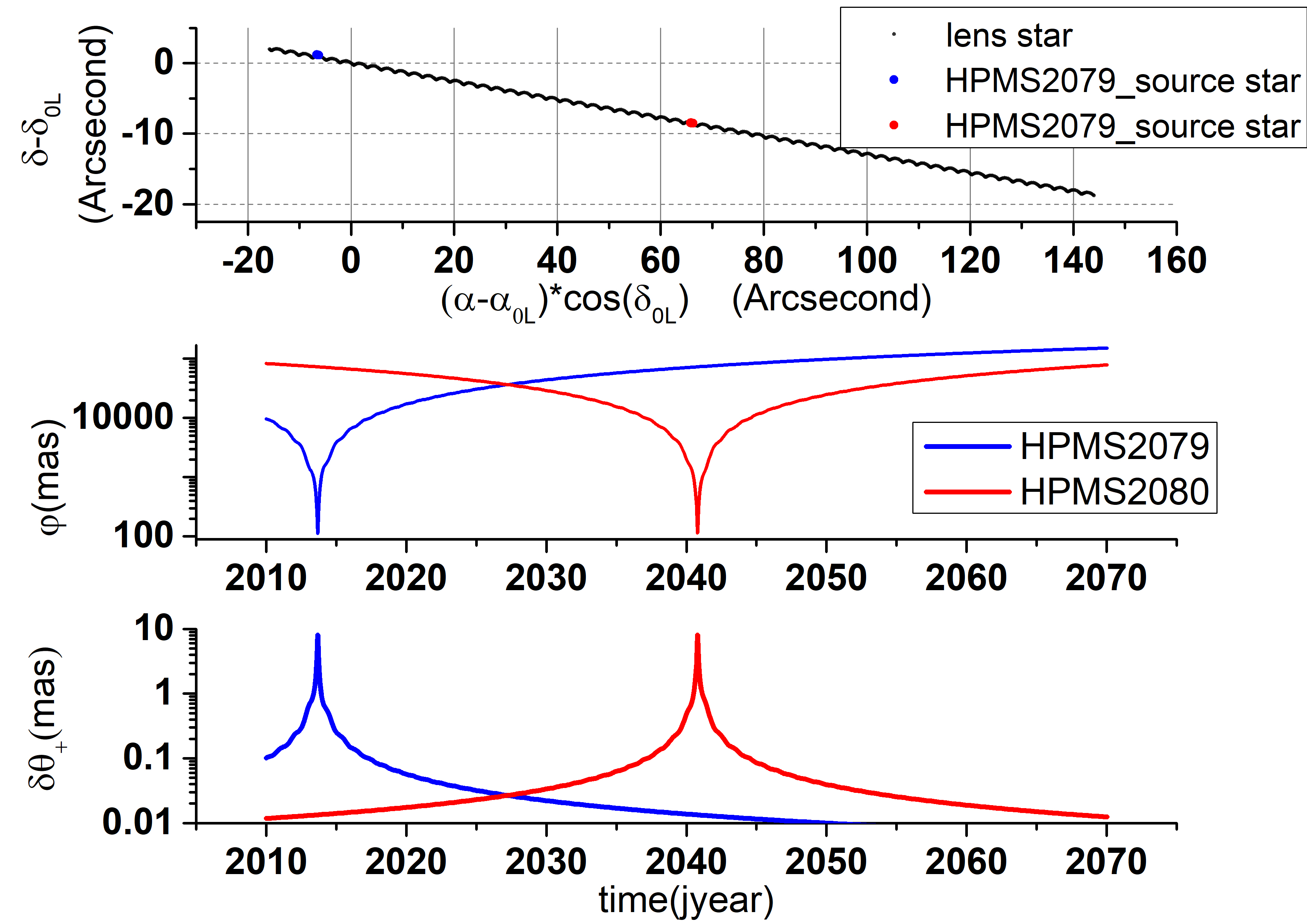}
	\caption{The data of the events HPMS2079 and HPMS2080. Top panel: the tracks of the lens star and the background star for two events in J2010-J2070. The black dots are the tracks of the lens star, the blue dots and the red dots are the tracks of the background star of the events HPMS2079 and HPMS2080 respectively. Middle panel: the change of the $\varphi$ of two events with time. The blue curve and the red curve represent the events HPMS2079 and HPMS2080 respectively. Bottom panel: the change of $\delta\theta_{+}$ with time, the blue curve and the red curve represent the events HPMS2079 and HPMS2080, respectively.}
	\label{Fig_HPMS2079,HPMS2080}
\end{figure}

(1)  71 astrometric microlensing events caused by 61 CYG A for Gaia soruce\_id 187204660934556480

In Gaia DR3, the lens star 61 CYG A will cause 71 astrometric microlensing events, as shown in Figure \ref{Fig_background78} and Figure \ref{Fig_lens2}, and the detailed data is listed in the Appendix (event names: HPMS4708 - HPMS4783). The lens star is the main sequence star, its spectral type is K, taken from Gaia DR3 astrophysical parameters. It is a  binary stars,  and has the annual parallax of 286mas and the proper motion of 5282mas / yr. We compare our results with those of \citet{Kluter22}. In 26 events searched by \citet{Kluter22}, one event is not included in our results, because the background star cannot match Gaia DR2 and external catalogues (see section \ref{sec:Checking for microlensing events}). The mass of the lens star is estimated by the table "Astrophysical parameters"  to be $(0.680 \pm 0.04) M_{\odot}$, which is slightly larger than that of \citet{Kluter22} ($0.621\pm 0.06) M_{\odot}$. There are 3 events with $\delta\theta_{+}>1mas$. It is noted from Figure \ref{Fig_background78} that the distance between the background stars and the proper motion track of the lens star exceeds $10^{\prime\prime}$ in about $45\%$ of the events, and finding that these events can increase the completeness of the predicted events. It should be emphasized that events with the distance between the background stars and the proper motion track of the lens star below $7^{\prime\prime}$ are predicted by \citet{Kluter22} except 3 events occurred before 2010 or after 2068.

(2) 52 astrometric microlensing events caused by LAWD 37 for Gaia source\_id 5332606522595645952

In Gaia DR3, the lens star LAWD 37 will cause 52 astrometric microlensing events, as shown in Figure \ref{Fig_background77} and Figure \ref{Fig_lens1}. The detailed data is shown in the appendix (event names: HPMS2109 - HPMS2132). The lens star is the white dwarf star, its spectral type is B, and it has the annual parallax of 216mas and the proper motion of 2684 mas / yr, taken from Gaia DR3.  \citet{Kluter22} had searched for 36 events, where 7 events are not included in our results, because the "astrometric fidelity" parameter value of the background star is less than 0.7, which do not meet the conditions in Table \ref{conditions_background}. We originally estimated the mass  of the lens star as $(0.77\pm 0.01) M_{\odot}$ from the white dwarf catalogue (\citet{Gentile2021}) labeled as 'WDJ114542.92-645029.46'.The mass of the lens star measured by \citet{McGill2023} is $(0.56\pm 0.08) M_{\odot}$, so we update the mass estimation and delete 23 events with $\delta\theta_{+} <0.1 mas $ due to mass changes. While \citet{Kluter22} estimated the mass of the lens star as $(0.65\pm 0.15) M_{\odot}$. For 10 events, the distance between the background stars and the proper motion track of the lens star exceeds $8^{\prime\prime}$. There are 4 events with $\delta\theta_{+}>1mas$, where 2 events (HPMS2079 and HPMS2080) have $\delta\theta_{+}>8mas$, and the minimum separation of star pairs is greater than 105mas, as shown in Figure \ref{Fig_lens1} and \ref{Fig_HPMS2079,HPMS2080}. Events with the distance between the background stars and the proper motion track of the lens star below $7^{\prime\prime}$ are predicted by \citet{Kluter22} except 4 events occurred before 2010 or after 2068.

As seen from the above two examples, the events searched by us and \citet{Kluter22} are different, because the different constraints for star pairs and mass estimation for lens star are used, specially we use the cone search method to find more background stars. 
It is also seen from Figure \ref{Fig_all_distance} that there are a small number of event with large distance of background stars from the proper motion track of the lens star.  Therefore, we suggest that our cone search method is helpful to find more events. It is noted that two lens stars will cause multiple events due to their high proper motion and parallax, the detection of many events caused by one lens will help us to improve lens mass accuracy.

\subsection{ The events with large astrometric signals ($\delta\theta_{+}$)}
\label{sec Events with large }

From the discussion in section \ref{sec:Basic facts}, it can be seen that only when $\theta_{sep}$ is large, for example, $\theta_{sep}>103mas$ for Gaia with FWHM=103mas \citep{Fabricius2016}, the signal $\delta\theta_{+}$ can be detected. Therefore, the events with large $\delta\theta_{+}$  discussed in this section should meet the condition of $\theta_{sep}>103mas$. We find that 348 events have $\delta\theta_{+}>0.5mas$, including 97 events with $\delta\theta_{+}>1mas$ and 2 events with $\delta\theta_{+}>8 mas$ discussed in Section \ref{sec Multiple astrometric microlensing events caused by a single lens star}. The masses of the lens stars in these events are concentrated in $0.4M_{\odot}<M_L<1 M_{\odot}$. However, we have not new predicted events with $\delta\theta_{+}>1 mas$ and $\theta_{sep}>103mas$.

\section{Summary and Conclusions}
\label{sec Summary and Conclusions}
In this paper, we select about 820000 potential lens stars from Gaia DR3 data for three types of stars, HPMS, NS and HMS, and obtain them to be about 470000, 160000 and 190000, respectively. Based on the cone search method, about 2260000 star pairs are initially selected, including 1240000, 260000 and 760000 star pairs with lens stars of HPMS, NS and HMS respectively. The relative motion of the star pair is approximated to be linear motion (not considering parallax). The mass of the lens star is estimated according to the "Astrophysical parameters", the White Dwarf Catalog (\citet{Gentile2021}) or the mass-luminosity relations, and then the minimum angular distance of the star pair in the period of J2010.0-J2070.0 is estimated. These star pairs are further selected to obtain 27057 star pairs, where HPMS, NS and HMS are 19401,3742 and 3914 respectively. For these star pairs, we search for their minimum angular distance to reserve the star pairs with $\delta\theta_{+}>0.1mas$ and remove unqualified star pairs according to Table \ref{conditions_background} and the equations (\ref{d_parallax}) and (\ref{cut pm}). We finally get 4500 events caused by 3558 lens stars, where 4279 events are caused by HPMS (their  background stars with two-, five- and six-parameter solutions are 907, 1453 and 1919 respectively), 220 events are caused by NS (their  background stars with two-, five- and six-parameter solutions are 71, 50 and 99 respectively), and only one event is caused by HMS (its background star has a two-parameter solution). There are 293 lens stars that can cause two or more events, where 5 lens stars will cause more than 50 events. Specially two lens stars (61 CYG A and  LAWD 37) are found to cause 71 and 52 events respectively. It is noted that two events, HPMS2079 and HPMS2080, have $\delta\theta_{+}>8 mas$ and $\theta_{sep}>103mas$.

Because we use the different conditions of lens star, background star, binary stars or co-moving stars, mass estimation of lens stars and search scopes of star pairs, the events we searched are not the same with ones searched by \citet{Kluter22}. We obtain 2836 events that are the same with those of \citet{Kluter22}, and 1664 new predicted event.

In the events we searched, 116 events have the distances of background stars from the proper motion path of lens stars more than $8^{\prime\prime}$ in the reference epoch, where the maximum distance is $16.6^{\prime\prime}$. Although $\delta\theta_{+}$ does not exceed 0.2 mas for 116 events, they increase the completeness of the predicted events. Therefore, the cone search method we used can help us to find more events. In addition, we have not found the event with lens star mass distributed in $3.2M_{\odot}<M_L<5M_{\odot}$, we consider to reduce the threshold of mass for HMS in future work and expect to find more events.

In the future, Gaia DR4 will provide the epoch astrometry\footnote{\url{https://www.cosmos.esa.int/web/gaia/newsletter/contents}}, which can be used to measure the shift of a handful events. The predicted events are expected to be followed up using space devices such as HST, JWST or Chinese Space Station Telescope (CSST) that will be launched at the near future \citep{CSST2018,CSST2019} .

\section*{Acknowledgements}

We thank the referee for useful comments and suggestions to improve the paper. This research work is financially supported by the National Natural Science Foundation of China (Grant Nos. 11403101, 12173085), Training Object Project of technological innovation talents in Yunnan Province(No. 202305AD160004) and the China Manned Space Project. We use the data from the European Space Agency (ESA) mission Gaia (https://www.cosmos.esa.int/gaia) and the Gaia Data Processing and Analysis Consortium (DPAC, https://www.cosmos.esa.int/web/gaia/dpac/consortium). Funding for the DPAC has been provided by national institutions, in particular the institutions participating in the Gaia Multilateral Agreement.

This research has made use of the data of ESASky, developed by the ESAC Science Data Centre (ESDC) team and maintained alongside other ESA science mission's archives at ESA's European Space Astronomy Centre (ESAC, Madrid, Spain)\citep{ESASky2017,ESASky2018}, the images of the DESI Legacy Imaging Surveys consisting of three individual and complementary projects: the Dark Energy Camera Legacy Survey (DECaLS), the Beijing-Arizona Sky Survey (BASS), and the Mayall z-band Legacy Survey (MzLS) \citep{LegacySurveys}, and the images of SkyMapper \citep{SkyMapper2016,SkyMapper2018,SkyMapper2019}.

\section*{Data Availability}

The data underlying this article are available in the article and in its spreadsheet available \href{https://osf.io/92p8r/?view\_only=178c791217c644aebc8c125ef0347b92}{online}.



\bibliographystyle{mnras}
\bibliography{reference} 

\begin{thebibliography}{}
\makeatletter
\relax
\def\mn@urlcharsother{\let\do\@makeother \do\$\do\&\do\#\do\^\do\_\do\%\do\~}
\def\mn@doi{\begingroup\mn@urlcharsother \@ifnextchar [ {\mn@doi@}
  {\mn@doi@[]}}
\def\mn@doi@[#1]#2{\def\@tempa{#1}\ifx\@tempa\@empty \href
  {http://dx.doi.org/#2} {doi:#2}\else \href {http://dx.doi.org/#2} {#1}\fi
  \endgroup}
\def\mn@eprint#1#2{\mn@eprint@#1:#2::\@nil}
\def\mn@eprint@arXiv#1{\href {http://arxiv.org/abs/#1} {{\tt arXiv:#1}}}
\def\mn@eprint@dblp#1{\href {http://dblp.uni-trier.de/rec/bibtex/#1.xml}
  {dblp:#1}}
\def\mn@eprint@#1:#2:#3:#4\@nil{\def\@tempa {#1}\def\@tempb {#2}\def\@tempc
  {#3}\ifx \@tempc \@empty \let \@tempc \@tempb \let \@tempb \@tempa \fi \ifx
  \@tempb \@empty \def\@tempb {arXiv}\fi \@ifundefined
  {mn@eprint@\@tempb}{\@tempb:\@tempc}{\expandafter \expandafter \csname
  mn@eprint@\@tempb\endcsname \expandafter{\@tempc}}}

\bibitem[\protect\citeauthoryear{Albareti et~al.,}{Albareti
  et~al.}{2017}]{SDSSDR13}
Albareti F.~D.,  et~al., 2017, \mn@doi [ApJS] {10.3847/1538-4365/aa8992}, 233

\bibitem[\protect\citeauthoryear{{Astropy Collaboration} et~al.,}{{Astropy
  Collaboration} et~al.}{2013}]{AstropyCollaboration}
{Astropy Collaboration} et~al., 2013, \mn@doi [\aap]
  {10.1051/0004-6361/201322068}, \href
  {https://ui.adsabs.harvard.edu/abs/2013A&A...558A..33A} {558, A33}

\bibitem[\protect\citeauthoryear{{Astropy Collaboration} et~al.,}{{Astropy
  Collaboration} et~al.}{2018}]{AstropyCollaboration18}
{Astropy Collaboration} et~al., 2018, \mn@doi [\aj] {10.3847/1538-3881/aabc4f},
  \href {https://ui.adsabs.harvard.edu/abs/2018AJ....156..123A} {156, 123}

\bibitem[\protect\citeauthoryear{Baines et~al.,}{Baines
  et~al.}{2017}]{ESASky2017}
Baines D.,  et~al., 2017, \mn@doi [Publications of the Astronomical Society of
  the Pacific] {10.1088/1538-3873/129/972/028001}, 129, 028001

\bibitem[\protect\citeauthoryear{Bellini, Anderson  \& Bedin}{Bellini
  et~al.}{2011}]{HST}
Bellini A.,  Anderson J.,   Bedin L.~R.,  2011, \mn@doi [PASP]
  {10.1086/659878}, 123, 622

\bibitem[\protect\citeauthoryear{{Belokurov} \& {Evans}}{{Belokurov} \&
  {Evans}}{2002}]{Belokurov}
{Belokurov} V.~A.,  {Evans} N.~W.,  2002, \mn@doi [\mnras]
  {10.1046/j.1365-8711.2002.05222.x}, \href
  {https://ui.adsabs.harvard.edu/abs/2002MNRAS.331..649B} {331, 649}

\bibitem[\protect\citeauthoryear{{Bond} et~al.,}{{Bond} et~al.}{2001}]{Bond}
{Bond} I.~A.,  et~al., 2001, \mn@doi [\mnras]
  {10.1046/j.1365-8711.2001.04776.x}, \href
  {https://ui.adsabs.harvard.edu/abs/2001MNRAS.327..868B} {327, 868}

\bibitem[\protect\citeauthoryear{{Bramich}}{{Bramich}}{2018}]{Bramich18b}
{Bramich} D.~M.,  2018, \mn@doi [\aap] {10.1051/0004-6361/201833505}, \href
  {https://ui.adsabs.harvard.edu/abs/2018A&A...618A..44B} {618, A44}

\bibitem[\protect\citeauthoryear{{Bramich} \& {Nielsen}}{{Bramich} \&
  {Nielsen}}{2018}]{Bramich18a}
{Bramich} D.~M.,  {Nielsen} M.~B.,  2018, \mn@doi [\actaa]
  {10.32023/0001-5237/68.3.1}, \href
  {https://ui.adsabs.harvard.edu/abs/2018AcA....68..183B} {68, 183}

\bibitem[\protect\citeauthoryear{Cao et~al.,}{Cao et~al.}{2018}]{CSST2018}
Cao Y.,  et~al., 2018, \mn@doi [MNRAS] {10.1093/mnras/sty1980}, 480, 2178

\bibitem[\protect\citeauthoryear{Collaboration et~al.,}{Collaboration
  et~al.}{2022}]{AstropyCollaboration2022}
Collaboration A.,  et~al., 2022, \mn@doi [The Astrophysical Journal]
  {10.3847/1538-4357/ac7c74}, 935, 167

\bibitem[\protect\citeauthoryear{Collaboration et~al.,}{Collaboration
  et~al.}{2023a}]{LegacySurveys}
Collaboration D.,  et~al., 2023a, Validation of the Scientific Program for the
  Dark Energy Spectroscopic Instrument, \mn@doi{10.48550/arXiv.2306.06307},
  \url {https://ui.adsabs.harvard.edu/abs/2023arXiv230606307D}

\bibitem[\protect\citeauthoryear{Collaboration et~al.,}{Collaboration
  et~al.}{2023b}]{Collaboration22}
Collaboration G.,  et~al., 2023b, \mn@doi [\aap] {10.1051/0004-6361/202243940},
  674, A1

\bibitem[\protect\citeauthoryear{Creevey et~al.,}{Creevey
  et~al.}{2023}]{Creevey2022}
Creevey O.~L.,  et~al., 2023, \mn@doi [\aap] {10.1051/0004-6361/202243688},
  674, A26

\bibitem[\protect\citeauthoryear{{Dominik} \& {Sahu}}{{Dominik} \&
  {Sahu}}{2000}]{Dominik}
{Dominik} M.,  {Sahu} K.~C.,  2000, \mn@doi [\apj] {10.1086/308716}, \href
  {https://ui.adsabs.harvard.edu/abs/2000ApJ...534..213D} {534, 213}

\bibitem[\protect\citeauthoryear{El-Badry, Rix  \& Heintz}{El-Badry
  et~al.}{2021}]{El-Badry2021}
El-Badry K.,  Rix H.-W.,   Heintz T.~M.,  2021, \mn@doi [MNRAS]
  {10.1093/mnras/stab323}, 506, 2269

\bibitem[\protect\citeauthoryear{Fabricius et~al.,}{Fabricius
  et~al.}{2016}]{Fabricius2016}
Fabricius C.,  et~al., 2016, \mn@doi [\aap] {10.1051/0004-6361/201628643}, 595,
  A3

\bibitem[\protect\citeauthoryear{Fabricius et~al.,}{Fabricius
  et~al.}{2021}]{Fabricius2021}
Fabricius C.,  et~al., 2021, \mn@doi [\aap] {10.1051/0004-6361/202039834}, 649,
  A5

\bibitem[\protect\citeauthoryear{Flewelling et~al.,}{Flewelling
  et~al.}{2020}]{Pan-STARRS1}
Flewelling H.~A.,  et~al., 2020, \mn@doi [ApJS] {10.3847/1538-4365/abb82d}, 251

\bibitem[\protect\citeauthoryear{Fouesneau et~al.,}{Fouesneau
  et~al.}{2023}]{Fouesneau2022}
Fouesneau M.,  et~al., 2023, \mn@doi [\aap] {10.1051/0004-6361/202243919}, 674,
  A28

\bibitem[\protect\citeauthoryear{{Gaia Collaboration} et~al.,}{{Gaia
  Collaboration} et~al.}{2016}]{Collaboration}
{Gaia Collaboration} et~al., 2016, \mn@doi [\aap]
  {10.1051/0004-6361/201629272}, \href
  {https://ui.adsabs.harvard.edu/abs/2016A&A...595A...1G} {595, A1}

\bibitem[\protect\citeauthoryear{Gardner et~al.,}{Gardner et~al.}{2023}]{JWST}
Gardner J.~P.,  et~al., 2023, \mn@doi [PASP] {10.1088/1538-3873/acd1b5}, 135,
  068001

\bibitem[\protect\citeauthoryear{Gentile~Fusillo et~al.,}{Gentile~Fusillo
  et~al.}{2021}]{Gentile2021}
Gentile~Fusillo N.~P.,  et~al., 2021, \mn@doi [MNRAS] {10.1093/mnras/stab2672},
  508, 3877

\bibitem[\protect\citeauthoryear{Giordano et~al.,}{Giordano
  et~al.}{2018}]{ESASky2018}
Giordano F.,  et~al., 2018, \mn@doi [Astronomy and Computing]
  {10.1016/j.ascom.2018.05.002}, 24, 97

\bibitem[\protect\citeauthoryear{Gong et~al.,}{Gong et~al.}{2019}]{CSST2019}
Gong Y.,  et~al., 2019, \mn@doi [AJ] {10.3847/1538-4357/ab391e}, 883, 203

\bibitem[\protect\citeauthoryear{Henden, Templeton, Terrell, Smith, Levine  \&
  Welch}{Henden et~al.}{2016}]{APASSDR9}
Henden A.~A.,  Templeton M.,  Terrell D.,  Smith T.~C.,  Levine S.,   Welch D.,
   2016, VizieR Online Data Catalog, p. II/336

\bibitem[\protect\citeauthoryear{{Hog}, {Novikov}  \& {Polnarev}}{{Hog}
  et~al.}{1995}]{Hog1995}
{Hog} E.,  {Novikov} I.~D.,   {Polnarev} A.~G.,  1995, \aap, \href
  {https://ui.adsabs.harvard.edu/abs/1995A&A...294..287H} {294, 287}

\bibitem[\protect\citeauthoryear{Hog et~al.,}{Hog et~al.}{2000}]{Tycho2}
Hog E.,  et~al., 2000, Astronomy and Astrophysics, 355, L27

\bibitem[\protect\citeauthoryear{Jablonska, Wyrzykowski, Rybicki, Kruszynska,
  Kaczmarek  \& Penoyre}{Jablonska et~al.}{2022}]{Jablonska2022}
Jablonska M.,  Wyrzykowski L.,  Rybicki K.~A.,  Kruszynska K.,  Kaczmarek Z.,
  Penoyre Z.,  2022, \mn@doi [\aap] {10.1051/0004-6361/202244656}, 666, L16

\bibitem[\protect\citeauthoryear{Kains et~al.,}{Kains et~al.}{2017}]{Kains2017}
Kains N.,  et~al., 2017, \mn@doi [AJ] {10.3847/1538-4357/aa78eb}, 843, 145

\bibitem[\protect\citeauthoryear{{Kl{\"u}ter}, {Bastian}, {Demleitner}  \&
  {Wambsganss}}{{Kl{\"u}ter} et~al.}{2018a}]{Kluter18a}
{Kl{\"u}ter} J.,  {Bastian} U.,  {Demleitner} M.,   {Wambsganss} J.,  2018a,
  \mn@doi [\aap] {10.1051/0004-6361/201833461}, \href
  {https://ui.adsabs.harvard.edu/abs/2018A&A...615L..11K} {615, L11}

\bibitem[\protect\citeauthoryear{{Kl{\"u}ter}, {Bastian}, {Demleitner}  \&
  {Wambsganss}}{{Kl{\"u}ter} et~al.}{2018b}]{Kluter18b}
{Kl{\"u}ter} J.,  {Bastian} U.,  {Demleitner} M.,   {Wambsganss} J.,  2018b,
  \mn@doi [\aap] {10.1051/0004-6361/201833978}, \href
  {https://ui.adsabs.harvard.edu/abs/2018A&A...620A.175K} {620, A175}

\bibitem[\protect\citeauthoryear{{Kl{\"u}ter}, {Bastian}, {Demleitner}  \&
  {Wambsganss}}{{Kl{\"u}ter} et~al.}{2022}]{Kluter22}
{Kl{\"u}ter} J.,  {Bastian} U.,  {Demleitner} M.,   {Wambsganss} J.,  2022,
  \mn@doi [\aj] {10.3847/1538-3881/ac4fc0}, \href
  {https://ui.adsabs.harvard.edu/abs/2022AJ....163..176K} {163, 176}

\bibitem[\protect\citeauthoryear{Kunder et~al.,}{Kunder et~al.}{2017}]{RAVEDR5}
Kunder A.,  et~al., 2017, \mn@doi [AJ] {10.3847/1538-3881/153/2/75}, 153, 75

\bibitem[\protect\citeauthoryear{Lam et~al.,}{Lam et~al.}{2022}]{Lam2022}
Lam C.~Y.,  et~al., 2022, \mn@doi [AJ] {10.3847/2041-8213/ac7442}, 933, L23

\bibitem[\protect\citeauthoryear{Lasker et~al.,}{Lasker et~al.}{2008}]{GSC2.3}
Lasker B.~M.,  et~al., 2008, \mn@doi [AJ] {10.1088/0004-6256/136/2/735}, 136,
  735

\bibitem[\protect\citeauthoryear{Lindegren et~al.,}{Lindegren
  et~al.}{2021}]{Lindegren2021}
Lindegren L.,  et~al., 2021, \mn@doi [\aap] {10.1051/0004-6361/202039709}, 649

\bibitem[\protect\citeauthoryear{Luberto, Martin, McGill, Leauthaud, Skemer  \&
  Lu}{Luberto et~al.}{2022}]{RN22}
Luberto J.,  Martin E.~C.,  McGill P.,  Leauthaud A.,  Skemer A.~J.,   Lu
  J.~R.,  2022, \mn@doi [AJ] {10.3847/1538-3881/ac9a41}, 164, 253

\bibitem[\protect\citeauthoryear{Mainzer et~al.,}{Mainzer
  et~al.}{2011}]{AllWISE}
Mainzer A.,  et~al., 2011, \mn@doi [AJ] {10.1088/0004-637x/731/1/53}, 731, 53

\bibitem[\protect\citeauthoryear{{Marrese}, {Marinoni}, {Fabrizio}  \&
  {Giuffrida}}{{Marrese} et~al.}{2017}]{Marrese2017}
{Marrese} P.~M.,  {Marinoni} S.,  {Fabrizio} M.,   {Giuffrida} G.,  2017,
  \mn@doi [\aap] {10.1051/0004-6361/201730965}, \href
  {https://ui.adsabs.harvard.edu/abs/2017A&A...607A.105M} {607, A105}

\bibitem[\protect\citeauthoryear{{Marrese}, {Marinoni}, {Fabrizio}  \&
  {Altavilla}}{{Marrese} et~al.}{2019}]{Marrese2019}
{Marrese} P.~M.,  {Marinoni} S.,  {Fabrizio} M.,   {Altavilla} G.,  2019,
  \mn@doi [\aap] {10.1051/0004-6361/201834142}, \href
  {https://ui.adsabs.harvard.edu/abs/2019A&A...621A.144M} {621, A144}

\bibitem[\protect\citeauthoryear{{McGill}, {Smith}, {Evans}, {Belokurov}  \&
  {Smart}}{{McGill} et~al.}{2018}]{McGill}
{McGill} P.,  {Smith} L.~C.,  {Evans} N.~W.,  {Belokurov} V.,   {Smart} R.~L.,
  2018, \mn@doi [\mnras] {10.1093/mnrasl/sly066}, \href
  {https://ui.adsabs.harvard.edu/abs/2018MNRAS.478L..29M} {478, L29}

\bibitem[\protect\citeauthoryear{McGill, Smith, Evans, Belokurov  \&
  Zhang}{McGill et~al.}{2019a}]{McGill2019}
McGill P.,  Smith L.~C.,  Evans N.~W.,  Belokurov V.,   Zhang Z.~H.,  2019a,
  \mn@doi [MNRAS] {10.1093/mnras/sty3344}, 483, 4210

\bibitem[\protect\citeauthoryear{{McGill}, {Smith}, {Evans}, {Belokurov}  \&
  {Lucas}}{{McGill} et~al.}{2019b}]{McGill19}
{McGill} P.,  {Smith} L.~C.,  {Evans} N.~W.,  {Belokurov} V.,   {Lucas} P.~W.,
  2019b, \mn@doi [\mnras] {10.1093/mnrasl/slz073}, \href
  {https://ui.adsabs.harvard.edu/abs/2019MNRAS.487L...7M} {487, L7}

\bibitem[\protect\citeauthoryear{McGill, Everall, Boubert  \& Smith}{McGill
  et~al.}{2020}]{McGill2020}
McGill P.,  Everall A.,  Boubert D.,   Smith L.~C.,  2020, \mn@doi [MNRAS]
  {10.1093/mnrasl/slaa118}, 498, L6

\bibitem[\protect\citeauthoryear{McGill et~al.,}{McGill
  et~al.}{2023}]{McGill2023}
McGill P.,  et~al., 2023, \mn@doi [MNRAS] {10.1093/mnras/stac3532}, 520, 259

\bibitem[\protect\citeauthoryear{{Miyamoto} \& {Yoshii}}{{Miyamoto} \&
  {Yoshii}}{1995}]{Miyamoto1995}
{Miyamoto} M.,  {Yoshii} Y.,  1995, \mn@doi [\aj] {10.1086/117616}, \href
  {https://ui.adsabs.harvard.edu/abs/1995AJ....110.1427M} {110, 1427}

\bibitem[\protect\citeauthoryear{Mustill, Davies  \& Lindegren}{Mustill
  et~al.}{2018}]{Mustill2018}
Mustill A.~J.,  Davies M.~B.,   Lindegren L.,  2018, \mn@doi [\aap]
  {10.1051/0004-6361/201833527}, 617

\bibitem[\protect\citeauthoryear{Ofek}{Ofek}{2018}]{Ofek2018}
Ofek E.~O.,  2018, \mn@doi [AJ] {10.3847/1538-4357/aadfeb}, 866, 144

\bibitem[\protect\citeauthoryear{Onken et~al.,}{Onken
  et~al.}{2019a}]{SkyMapperDR2}
Onken C.~A.,  et~al., 2019a, \mn@doi [PASA] {10.1017/pasa.2019.27}, 36

\bibitem[\protect\citeauthoryear{Onken et~al.,}{Onken
  et~al.}{2019b}]{SkyMapper2019}
Onken C.~A.,  et~al., 2019b, \mn@doi [Publications of the Astronomical Society
  of Australia] {10.1017/pasa.2019.27}, 36, e033

\bibitem[\protect\citeauthoryear{{Paczynski}}{{Paczynski}}{1986}]{Paczynski}
{Paczynski} B.,  1986, \mn@doi [\apj] {10.1086/163919}, \href
  {https://ui.adsabs.harvard.edu/abs/1986ApJ...301..503P} {301, 503}

\bibitem[\protect\citeauthoryear{{Paczynski}}{{Paczynski}}{1995}]{Paczynski95}
{Paczynski} B.,  1995, \actaa, \href
  {https://ui.adsabs.harvard.edu/abs/1995AcA....45..345P} {45, 345}

\bibitem[\protect\citeauthoryear{{Paczynski}}{{Paczynski}}{1996}]{Paczynski96}
{Paczynski} B.,  1996, \mn@doi [\araa] {10.1146/annurev.astro.34.1.419}, \href
  {https://ui.adsabs.harvard.edu/abs/1996ARA&A..34..419P} {34, 419}

\bibitem[\protect\citeauthoryear{Rybizki et~al.,}{Rybizki
  et~al.}{2022}]{Rybizki22}
Rybizki J.,  et~al., 2022, \mn@doi [MNRAS] {10.1093/mnras/stab3588}, 510, 2597

\bibitem[\protect\citeauthoryear{{Sahu} et~al.,}{{Sahu} et~al.}{2017}]{Sahu17}
{Sahu} K.~C.,  et~al., 2017, \mn@doi [Science] {10.1126/science.aal2879}, \href
  {https://ui.adsabs.harvard.edu/abs/2017Sci...356.1046S} {356, 1046}

\bibitem[\protect\citeauthoryear{{Sahu} et~al.,}{{Sahu} et~al.}{2022}]{Sahu}
{Sahu} K.~C.,  et~al., 2022, \mn@doi [\apj] {10.3847/1538-4357/ac739e}, \href
  {https://ui.adsabs.harvard.edu/abs/2022ApJ...933...83S} {933, 83}

\bibitem[\protect\citeauthoryear{Skrutskie et~al.,}{Skrutskie
  et~al.}{2006}]{2MASS}
Skrutskie M.~F.,  et~al., 2006, \mn@doi [AJ] {10.1086/498708}, 131, 1163

\bibitem[\protect\citeauthoryear{{Smith} et~al.,}{{Smith}
  et~al.}{2018}]{Smith2018}
{Smith} L.~C.,  et~al., 2018, \mn@doi [\mnras] {10.1093/mnras/stx2789}, \href
  {https://ui.adsabs.harvard.edu/abs/2018MNRAS.474.1826S} {474, 1826}

\bibitem[\protect\citeauthoryear{Steinmetz et~al.,}{Steinmetz
  et~al.}{2020}]{RAVEDR6}
Steinmetz M.,  et~al., 2020, \mn@doi [AJ] {10.3847/1538-3881/ab9ab8}, 160, 83

\bibitem[\protect\citeauthoryear{{Torra} et~al.,}{{Torra}
  et~al.}{2021}]{Torra2021}
{Torra} F.,  et~al., 2021, \mn@doi [\aap] {10.1051/0004-6361/202039637}, \href
  {https://ui.adsabs.harvard.edu/abs/2021A&A...649A..10T} {649, A10}

\bibitem[\protect\citeauthoryear{{Udalski}}{{Udalski}}{2003}]{Udalski}
{Udalski} A.,  2003, \actaa, \href
  {https://ui.adsabs.harvard.edu/abs/2003AcA....53..291U} {53, 291}

\bibitem[\protect\citeauthoryear{{Walker}}{{Walker}}{1995}]{Walker1995}
{Walker} M.~A.,  1995, \mn@doi [\apj] {10.1086/176367}, \href
  {https://ui.adsabs.harvard.edu/abs/1995ApJ...453...37W} {453, 37}

\bibitem[\protect\citeauthoryear{Wolf et~al.,}{Wolf
  et~al.}{2016}]{SkyMapper2016}
Wolf C.,  et~al., 2016, \mn@doi [SkyMapper Early Data Release]
  {10.4225/41/572ff2c5ebd30}

\bibitem[\protect\citeauthoryear{Wolf et~al.,}{Wolf
  et~al.}{2018}]{SkyMapper2018}
Wolf C.,  et~al., 2018, \mn@doi [Publications of the Astronomical Society of
  Australia] {10.1017/pasa.2018.5}, 35, e010

\bibitem[\protect\citeauthoryear{Wyrzykowski et~al.,}{Wyrzykowski
  et~al.}{2022}]{Wyrzykowski2022}
Wyrzykowski L.,  et~al., 2022, Gaia Data Release 3: Microlensing Events from
  All Over the Sky, \url
  {https://ui.adsabs.harvard.edu/abs/2022arXiv220606121W}

\bibitem[\protect\citeauthoryear{Zacharias et~al.,}{Zacharias
  et~al.}{2015}]{URAT1}
Zacharias N.,  et~al., 2015, \mn@doi [AJ] {10.1088/0004-6256/150/4/101}, 150

\bibitem[\protect\citeauthoryear{{Zurlo} et~al.,}{{Zurlo} et~al.}{2018}]{Zurlo}
{Zurlo} A.,  et~al., 2018, \mn@doi [\mnras] {10.1093/mnras/sty1805}, \href
  {https://ui.adsabs.harvard.edu/abs/2018MNRAS.480..236Z} {480, 236}

\bibitem[\protect\citeauthoryear{van Leeuwen}{van Leeuwen}{2007}]{Hipparcos-2}
van Leeuwen F.,  2007, \mn@doi [\aap] {10.1051/0004-6361:20078357}, 474, 653

\makeatother
\end{thebibliography}




\appendix


\bsp	
\label{lastpage}
\end{document}